\documentclass[12pt]{article}

\usepackage{latexsym,amsfonts,theorem}

\theoremstyle{change}
\newtheorem{theorem}{Theorem.}
\newtheorem{lemma}[theorem]{Lemma.}

\newtheorem{proposition}[theorem]{Proposition.}

\newcommand{\zerarcounters}
{
\setcounter{equation}{0}
\setcounter{theorem}{0}
}

\newcommand{\Fullbox}{\bigskip\hfill{\rule{2.5mm}{2.5mm}}}

\DeclareRobustCommand{\teb}[1]{
 \begingroup
  \mathchoice
    {\hbox{{$\displaystyle\oalign{$#1$\crcr\hidewidth\vbox 
             to.2ex{\hbox{\LARGE{\char126}}\vss}\hidewidth}  $}}}%
    {\hbox{{$\textstyle\oalign{$#1$\crcr\hidewidth\vbox 
             to.2ex{\hbox{\LARGE{\char126}}\vss}\hidewidth}  $}}}%
    {\hbox{{$\scriptstyle\oalign{$\scriptstyle #1$\crcr\hidewidth\vbox 
             to.2ex{\hbox{\char126}\vss}\hidewidth}  $}}}%
    {\hbox{{$\scriptscriptstyle
             \oalign{$\scriptscriptstyle #1$\crcr\hidewidth\vbox 
             to.2ex{\hbox{\char126}\vss}\hidewidth}  $}}}
 \endgroup}

\newcommand{\und}{\underline}

\newcommand{\undm}{\underline{m}}
\newcommand{\undn}{\underline{n}}

\newcommand{\undv}{\underline{v}}

\newcommand{\calB}{{\mathcal B}}
\newcommand{\calC}{{\mathcal C}}

\newcommand{\calE}{{\mathcal E}}

\newcommand{\calK}{{\mathcal K}}
\newcommand{\calL}{{\mathcal L}}

\newcommand{\calN}{{\mathcal N}}

\newcommand{\calQ}{{\mathcal Q}}
\newcommand{\calR}{{\mathcal R}}

\newcommand{\calU}{{\mathcal U}}

\newcommand{\be}{\begin{equation}}
\newcommand{\ee}{\end{equation}}
\newcommand{\bma}{\begin{displaymath}}
\newcommand{\ema}{\end{displaymath}}
\newcommand{\bc}{\begin{center}}
\newcommand{\ec}{\end{center}}

\newcommand{\bear}{\begin{eqnarray}}
\newcommand{\eear}{\end{eqnarray}}

\newcommand{\uflex}
{{\scriptstyle {\raise 9pt\hbox{$\backslash$}\,\!\!\!\!\!\Bigg\vert}}}

\newcommand{\Z}{\mathbb Z}
\newcommand{\N}{\mathbb N}
\newcommand{\R}{\mathbb R}
\newcommand{\C}{\mathbb C}

\newcommand{\UM}{{1\! \rm{l}}}

\newcommand{\eps}{\epsilon}

\newcommand{\EndofStatement}{\samepage\bigskip\hfill\Box}
\newcommand{\EndofProof}{\Fullbox}

\newcommand{\tebom}{ \teb{\omega}_f }
\newcommand{\undom}{\underline{\omega}}

\newcommand{\novomod}[1]{\ll\!\! #1 \!\!\gg}
\newcommand{\meufloor}[1]{\lfloor #1\rfloor}

\begin{document}

\thispagestyle{empty}

\begin{center}
\begin{Large}
  
  {\bf Converging Perturbative Solutions of the Schr\"odinger Equation for a
    Two-Level System with a Hamiltonian Depending Periodically on Time
    }

\end{Large}

\vspace{4cm}

     Jo\~ao C. A. Barata\footnote{Partially supported by CNPq.}

      \vspace{.2cm}

  Instituto de F\'{\i}sica 

  Universidade de S\~ao Paulo

  Caixa Postal 66 318

  05315 970. S\~ao Paulo. SP. Brasil

      \vspace{.3cm}

  E-mail: jbarata{@}fma.if.usp.br

\end{center}

\vspace{4cm}

{\it
  \noindent{\bf Abstract.} We study the Schr\"odinger equation of
  a class of two-level systems under the action of a periodic
  time-dependent external field in the situation where the energy
  difference $2\eps$ between the free energy levels is sufficiently
  small with respect to the strength of the external interaction.
  Under suitable conditions we show that this equation has a solution
  in terms of converging power series expansions in $\eps$.  In
  contrast to other expansion methods, like in the Dyson expansion,
  the method we present is not plagued by the presence of ``secular
  terms''. Due to this feature we were able to prove absolute and
  uniform convergence of the Fourier series involved in the
  computation of the wave functions and to prove absolute convergence
  of the $\eps$-expansions leading to the ``secular frequency'' and to
  the coefficients of the Fourier expansion of the wave function. }

\vspace{2cm}


\newpage

\tableofcontents

\newpage

\section{Introduction}
\label{Introduction}
\zerarcounters

Let us consider the following Hamiltonian for a two-level system under
the action of an external time-dependent field
\be
     H_1(t) \; = \; H_0 +H_I(t) \; = \; \eps \sigma_3 - f(t)\sigma_1 
\label{Hamiltoniano1}
\ee 
and the corresponding Schr\"odinger equation\footnote{For simplicity
  we shall adopt here a system of units with $\hbar =1$.}
\be
     i\partial_t \Psi(t) \; = \; H_1(t)\Psi(t) ,
\label{equacaodeSchroedingerparaPsi}
\ee 
with $\Psi: \R \to \C^2$.  Here $f(t)$ is a function of time $t$ and $
\eps\in\R$ is a parameter representing half of the energy difference
between the ``free'' (i.e., for $f\equiv 0 $) energy levels.  The
symbols $\sigma_1$, $ \sigma_2$ and $ \sigma_3$ denote the Pauli
matrices in their usual representations:
$$
\sigma_1 = \left(
                  \begin{array}{rr} 0 & 1 \\ 1 & 0\end{array}
           \right), \mbox{ } 
\sigma_2 = \left(
                  \begin{array}{rr} 0 & -i \\ i & 0\end{array}
           \right) \mbox{ and }
\sigma_3 = \left(
                  \begin{array}{rr} 1 & 0 \\ 0 & -1\end{array}
           \right),
$$
satisfying the commutation relations $[\sigma_1, \; \sigma_2] =
2i\sigma_3$, plus cyclic permutations.  

The ``interaction Hamiltonian'' $ H_I (t) := -f(t)\sigma_1$ represents
a time-dependent external interaction coupled to the system inducing
transitions between the two eigen-states of the free Hamiltonian $H_0
:= \eps \sigma_3$.  The situation where $ \eps$ is ``small''
characterizes the ``large coupling domain''
\cite{Walter1}-\cite{WreszinskiCasmeridis}.

The system described above is certainly one of the simplest
non-trivial time-depending quantum systems and the study of the
solutions of (\ref{equacaodeSchroedingerparaPsi}) is of basic
importance for many physical applications as, e.g., in quantum optics
or in problems of quantum tunnelling. 

Equation (\ref{equacaodeSchroedingerparaPsi}) has been analysed by
many authors in various approximations. In the wide literature on this
subject we mention the pioneering work of Autler and Townes
\cite{AutlerTownes}, where these authors studied the solutions of
(\ref{equacaodeSchroedingerparaPsi}) for the case where, in our
notation, $f(t) = -2\beta\cos (\omega t) $, $\beta \in \R$.  Their
work is exact but non-rigorous and involved a combination of the
method of continued fractions, for relating the coefficients the
Fourier decomposition of the wave functions, with numerical analysis.
No proof has been obtained that the continued fractions converge and
further unjustified restrictions have been made in order to transform
some transcendental equations into low order algebraic equations,
which are then solved either exactly or, specially for strong fields,
numerically.

For a recent review on the mathematical theory of quantum systems
submited to time-depending periodic and quasi-periodic perturbations
see \cite{Walter1}. For an introduction to the subject of ``quantum
chaos'' and quantum stability, see \cite{Jauslin}. See also
\cite{WreszinskiCasmeridis} for a spectral analysis of the
quasi-energy operator for two-level atoms in the quasi-periodic case.

In \cite{qp} we studied the system described by
(\ref{equacaodeSchroedingerparaPsi}) in the situation where $f$ is a
quasi-periodic function of time and a special sort of perturbative
expansion (power series expansion in $\eps$) has been developed. Its
main virtue is to be free of the so-called ``secular terms'', i.e.,
polynomials in $t$ that appear order by order in perturbation theory
and that spoil the analysis of convergence of the series and the
proofs of quasi-periodicity of the perturbative terms.  Although we
have not been able to prove convergence of our power series expansion
in the general case where $ f$ is quasi-periodic it has been
established that the coefficients of the expansion are indeed
quasi-periodic functions of time.

One of the obstacles found in the attempt to prove convergence of the
series is the presence of ``small denominators''. This typical feature
of perturbative approximations for solutions of differential equations
with quasi-periodic coefficients is well known as one of the main
sources of problems in the mathematically precise treatment of such
equations.

On what concerns proofs of convergence it should, therefore, be
expected that better results could be obtained if the function $f$
were restricted to be \underline{periodic} since, in this case, no
problems with small denominators should afflict our expansions.

However, the problem with small denominators is not the only problem
to be faced in the perturbative expansion of \cite{qp}. In this paper
we show how to circumvent the additional sources of difficulties and
to finally establish convergence of our perturbative expansion for
\underline{periodic} $f$. 

By a time-independent unitary transformation, representing a rotation
of $\pi/2$ around the 2-axis, we may replace $ H_1(t)$ by 
\be 
    H_2 (t) \; := \;
    \left(e^{-i\pi\sigma_2 /4}\right) \, H_1(t)\, 
    \left(e^{ i\pi\sigma_2 /4}\right) \; = \; \eps \sigma_1 +
    f(t)\sigma_3 
\ee 
and the Schr\"odinger equation becomes 
\be 
   i\partial_t \Phi(t) \; = \; H_2(t)\Phi(t),
    \label{schroedingerPhi}
\ee
with
\be
   \Phi (t) \; := \; e^{-i\pi\sigma_2 /4}\Psi(t) .
\ee

The theorem below, proven in \cite{qp}, presents the solution of the
Schr\"odinger equation (\ref{schroedingerPhi}) in terms of particular
solutions of a generalized Riccati equation.

\begin{theorem}  
Let $f: \R \to \R$, $f \in C^1 (\R)$ and $\eps \in \R$ and let 
$ g: \R \to \C$, $g \in C^1 (\R) $, be a particular solution of the
generalized Riccati equation
\be
  G' -i G^2 - 2 i f G + i \eps^2 = 0 .
\label{primeiraequacaodeRiccati}
\ee
Then, the function $\Phi: \R \to \C^2$ given by
\be
\displaystyle 
\Phi(t) \; = \; 
  \left(
    \begin{array}{c} 
       \phi_+ (t) \\ \phi_- (t)  
    \end{array} 
  \right) \; = \;  U(t)\Phi(0) \; = \;  U(t, \; 0)\Phi(0) , 
\label{solucaodaequacaodeSchroedinger}
\ee
where
\be
 U(t)  \; := \;
 \left(
  \begin{array}{cc}
     R(t)\left( 1 + i g(0) S(t)\right) & -i \eps R(t)S(t) \\
     &  \\
      -i \eps\overline{R(t)}\; \overline{S(t)} & 
                        \overline{R(t)}\left(
                         1 - i\;\overline{g(0)}\; \overline{S(t)}\right)
    \end{array}
   \right) ,
\label{definicaodeU}
\ee
with
\be
    R(t) \; := \; \exp\left( -i \int_0^t (f(\tau) + g(\tau))\,d\tau\right)
\label{definicaodeR}
\ee
and
\be
    S(t) \; := \; \int_0^t R(\tau )^{-2} \; d\tau 
\label{definicaodeS}
\ee
is a solution of the Schr\"odinger equation (\ref{schroedingerPhi})
with initial value
$
\displaystyle 
\Phi(0) \; = \; 
  \left(
    \begin{array}{c} 
       \phi_+ (0) \\ \phi_- (0)  
    \end{array} 
  \right)  \in \C^2
$.
$\EndofStatement  $
\label{proposicaosobreaformadasolucaoemtermosdeg}
\end{theorem}

For a proof of Theorem
\ref{proposicaosobreaformadasolucaoemtermosdeg}, see \cite{qp}.
Let us briefly describe some of the ideas leading to Theorem
\ref{proposicaosobreaformadasolucaoemtermosdeg} and to other results
of \cite{qp}.  As we saw in \cite{qp}, the solutions of the
Schr\"odinger equation (\ref{schroedingerPhi}) can be studied in terms
of the solutions of a particular complex version of Hill's equation:
\be 
        \phi''(t)  +  \left(  i f'(t) + \eps^2 +
         f(t)^2 \right)\phi (t) = 0 .
\label{Hillprimeira}
\ee
In fact, a simple computation (see \cite{qp}) shows that the
components $\phi_\pm$ of
$ 
\Phi (t) 
$
satisfy precisely
\be
\begin{array}{lcl}
\phi_+''  +  \left( + i f' + \eps^2 +
         f^2 \right)\phi_+  & = & 0 
\\
  & &
\\
\phi_-''  +  \left( - i f' + \eps^2 +
         f^2 \right)\phi_-  & = & 0 
\end{array} \;\;\;\; .
\label{Hillparapm}
\ee

As a side remark we note that equations (\ref{Hillparapm}) are simpler
and more convenient than the equations obtained by separating $\psi_+$
and $\psi_- $ from (\ref{equacaodeSchroedingerparaPsi}):
\be
\begin{array}{lcl}
\displaystyle
 \psi_+'' -  \left(\frac{f'}{f} \right)\psi_+' + \left(\eps^2 +f^2 -i\eps
 \left(\frac{f'}{f} \right) \right)\psi_+ & = & 0
\\
 & &
\\
\displaystyle
\psi_-'' -  \left(\frac{f'}{f} \right)\psi_-' + \left(\eps^2 +f^2 +i\eps
 \left(\frac{f'}{f} \right) \right)\psi_- & = & 0
\end{array} \;\;\;\; .
\ee
This last pair of equations, mentioned (but not used) 
in \cite{AutlerTownes}, is mathematically less
convenient because the coefficient $ f'/f$ can be discontinuous and
unbounded in typical cases as, for instance when $ f(t) = -2\beta
\cos(\omega t)$, the case analysed in \cite{AutlerTownes}.

In \cite{qp} we attempted to solve (\ref{Hillprimeira}) using the Ansatz 
\be
\phi (t) = \exp \left(-i\int_0^t (f(\tau)+g(\tau))
  d\tau\right) .
\label{AnsatzparasolucaodeHill}
\ee
It follows that $g$ has to satisfy the generalized Riccati equation
(\ref{primeiraequacaodeRiccati}) and we tried to find solutions for
$g$ in terms of a power expansion in $\eps$ like
\be
    g(t) \; = \; q(t)\sum_{n=1}^{\infty}\eps^n \,c_n(t), 
 \label{expansaodeg}
\ee
where
\be
       q(t) \; := \; \exp\left( i\int_0^t \, f(\tau)d\tau \right) .
\label{definicaodeq}
\ee

The heuristic idea behind the Ans\"atze (\ref{AnsatzparasolucaodeHill})
and (\ref{expansaodeg}) is the following. For $\eps \equiv 0$ a
solution for (\ref{Hillprimeira}) is given by $ \exp \left(-i\int_0^t
  f(\tau) d\tau\right)$. Thus, in (\ref{AnsatzparasolucaodeHill}) and
(\ref{expansaodeg}) we are searching for solutions in terms of an
``effective external field'' of the form $ f+g$, with $ g$ vanishing
for $\eps = 0$.

Notice that a solution of the form (\ref{AnsatzparasolucaodeHill})
leads to only one of the two independent solutions of the second order
Hill's equation (\ref{Hillprimeira}).  The complete solution of the
Schr\"odinger equation (\ref{schroedingerPhi}) in terms of solutions of
the generalized Riccati equation (\ref{primeiraequacaodeRiccati}) is
that described in Theorem
\ref{proposicaosobreaformadasolucaoemtermosdeg}.

As mentioned above, perturbative solutions of quasi-periodically
time-dependent systems are usually plagued by small denominators and
by the presence of the so-called ``secular terms''.  In \cite{qp} we
discovered a particular way to eliminate completely the secular terms
from the perturbative expansion of $g$ (see Appendix
\ref{ShortDescription}) and we were able to show, under some special
assumptions, that the coefficients $ c_n(t)$ are all quasi-periodic
functions. In \cite{qp} we proved convergence of our perturbative
solution in the somewhat trivial case where $ f(t)$ is a non-zero
constant function.  Unfortunately no conclusion could be drawn about
the convergence of the perturbative expansion for $g$ in the general
case of quasi-periodic $f$. We conjectured, however, that our
expansion is uniformly convergent at least in the situation where
$f(t) - M(f)$ is uniformly small. Here $ M(h)$ is the so-called ``mean
value'' of an {\it almost periodic} function $ h$, defined as (see,
e.g. \cite{Katznelson})
\be 
     M(h) \; := \; \lim_{T\to\infty}
     \frac{1}{2T}\int_{-T}^{T} \, h(t) \, dt .  
\ee

The technically central result of the present paper is the proof that,
under suitable assumptions, the series (\ref{expansaodeg}) converges
absolutely and uniformly on $\R$ as a function of time for $|\eps |$
small enough and $ f$ periodic. This is the content of Theorem
\ref{teoremasobredecaimentodosCns}. Moreover, we show that the
functions $ c_n$ and, hence, $g$, have uniformly converging Fourier
series representations. We use this fact together with the solution
(\ref{definicaodeU}) to find the Floquet representation of the
components $\phi_\pm $ of the wave function in terms of uniformly
converging Fourier series representations. This is the content of
Theorem \ref{Teoremaprincipalsobreopropagador}. Absolutely converging
power series in $ \eps$ for the Fourier coefficients and for the
secular frequency are also presented.

We believe that the methods employed in this paper are also of
importance for the general theory of Hill's equation. It would be of
great interest to know whether the ideas described in \cite{qp} and
here can be generalized and applied to a larger class of Hill's
equations than those we studied so far.

\subsection{The Main Result}

On what concerns the solutions of the Schr\"odinger equation
(\ref{schroedingerPhi}) the next theorem summarises our main
results.

\begin{theorem}
Let $ f$ be a real $T_\omega$-periodic function of time ($T_\omega :=
2\pi/\omega$) whose Fourier decomposition
\be
   f(t) \; = \;  \sum_{ n \in \Z }
   F_{n} e^{i n \omega t},
\label{primeiradecomposicaodeFourierdef}
\ee
with $\omega >0$, contains only a finite number of terms, i.e., the
set of integers $\{ n\in\Z | \; F_n \neq 0\}$ is a finite set.
Moreover, assume that $F_0 =0$.

Consider the two following mutually exclusive conditions on $f$:

I) $M(q^2)\neq 0$. 

II) $M(q^2) = 0$ but $M(\calQ_1) \neq 0$, where
\be
   \calQ_1(t) \; := \; q(t)^2  \int_0^t q^{-2}(\tau ) d\tau .
\label{definicaodecalQ}
\ee

Then, for each $ f$ as above, satisfying condition I or II, there
exists a constant $K >0$ (depending on the Fourier coefficients $\{
F_n , \; n\in \Z \; , n\neq 0\}$ and on $\omega>0$) such that, for
each $ \eps$ with $|\eps| < K$, there exist $\Omega \in \R$ and
$T_\omega$-periodic functions $u_{11}^\pm$ and $u_{12}^\pm$ such that
the propagator $ U(t)$ of (\ref{solucaodaequacaodeSchroedinger}) can
be written as
\be 
U(t) \; =
\; \left(
 \begin{array}{cc}
    U_{11}(t) & U_{12}(t) \\  & \\ U_{21}(t) & U_{22}(t) 
   \end{array}
\right)
\; = \;
\left(
 \begin{array}{cc}
    U_{11}(t) & U_{12}(t) \\ & \\ -\overline{U_{12}(t)} & \overline{U_{11}(t)} 
   \end{array}
\right) ,
\label{lisudfybvgwF}
\ee
with
\bear
U_{11}(t) & = & e^{-i\Omega t}\, u_{11}^-(t) + e^{i\Omega t}\, u_{11}^+ (t),
\label{U11}
\\
U_{12}(t) & = & e^{-i\Omega t}\, u_{12}^- (t) + e^{i\Omega t}\, u_{12}^+ (t).
\label{U12}
\eear
The functions  $u_{11}^\pm$ and $u_{12}^\pm$ have absolutely and 
uniformly converging Fourier expansions
$$
u_{11}^\pm(t) \; = \; \sum_{n\in \Z} \calU_{11}^{\pm}(n)e^{i n \omega t},
$$
$$
u_{12}^\pm(t) \; = \; \sum_{n\in \Z} \calU_{12}^{\pm}(n)e^{i n \omega t}.
$$
Moreover, under the same assumptions, $ \Omega$ and the Fourier coefficients
$ \calU_{11}^{\pm}(n)$ and $ \calU_{12}^{\pm}(n)$ can be expressed in
terms of absolutely converging power series on $ \eps$.
$\EndofStatement$
\label{Teoremaprincipalsobreopropagador}
\end{theorem}

\noindent {\bf Remarks on Theorem \ref{Teoremaprincipalsobreopropagador}}
\begin{enumerate}
\item Expressions (\ref{U11}) and (\ref{U12}) represent the so-called
  ``Floquet form'' of the matrix elements $U_{11}(t)$ and $U_{12}(t)$.
  The frequency $\Omega$ is called the ``secular frequency''.
\item In this paper we will assume that $F_0 = 0$. Results on the
  almost resonant case $F_0 \neq 0$, with $F_0/\omega$ satisfying some
  appropriated Diophantine conditions, will appear in a forthcoming
  publication \cite{JCABinprep}.
\item The physically realistic condition that the Fourier
  decomposition of $ f$ contains only a finite number of terms can be
  weakened. The only condition we use is the fast decay for $
  |m|\to\infty$ of the Fourier coefficients $ Q_m$ of the function $
  q(t)$ (defined in (\ref{definicaodeq})), as found in Proposition
  \ref{proposicaosobreodecaimentodeQm}.
\item The second equality in (\ref{lisudfybvgwF}) is due to
  (\ref{definicaodeU}). 
\item It is important to stress that conditions I and II are
  restrictions on the function $ f$ and not on the parameter $\eps$.
\item Possibly there are other conditions beyond {\it I} and {\it II}
  which could be considered, but they have not been explored so far.
  They are relevant in some cases.  Theorem
  \ref{Teoremaprincipalsobreopropagador} still does not provide a
  complete solution of (\ref{schroedingerPhi}) for all possible
  periodic functions $ f$, but examples and some qualitative arguments
  show that the remaining cases are rather exceptional. For instance,
  for $f(t) = \varphi_1\cos(\omega t) + \varphi_2\sin(\omega t)$
  condition I covers all pairs $(\varphi_1, \; \varphi_2)\in\R^2$,
  except only the countable family of circles centered at the origin
  with radius $ x_a \omega/2$, $ a = 1, \,2, \ldots$, where $ x_a$ if
  the $a$-th zero of $J_0$ in $ \R_+$ ($ J_0$ is the Bessel function
  of order zero). However, in these circles condition II is nowhere
  fulfilled.  See the discussion in Section \ref{hierarquias}.
\item From the computational point of view the solution given by our
  method can be easily implemented in numerical programs and has been
  successfully tested, providing ways to study our two-level system
  for large times with controllable errors (due to the uniform
  convergence).  Results on these numerical studies will be published
  elsewhere.
\item Unitarity of $U(t)$ for all $ t\in \R $ is a consequence of
  Dyson's expansion (see f.i. \cite{ReedSimonII}). 
\item Conditions I and II define, in principle, distinct solutions of
  the generalized Riccati equation (\ref{primeiraequacaodeRiccati})
  and, hence, of the Schr\"odinger equation (\ref{schroedingerPhi}).
  To fix a name we will call these solutions ``classes'' of solutions.

\end{enumerate}

\subsection{Remarks on the Notation}

Let us make some remarks on the notation we use here 
and recall the notation used
in \cite{qp}.  Given the Fourier representation\footnote{For
  convenience we adopt here a different notation as that found in
  \cite{qp}, where the Fourier decomposition of $ f$ was written as
$\displaystyle
 f(t) \; = \; \sum_{\teb{m}\in \Z^B} \, f_{\teb{m}} \, 
                                         e^{i\teb{m}\cdot \tebom t}
$.}
\be
 f(t) \;\; = \;\; \sum_{\teb{m}\in \Z^B} \; F_{\teb{m}} \; 
                                         e^{i\teb{m}\cdot \tebom t}
\label{representacaodefcomofuncaoquasi-periodica}
\ee
of the quasi-periodic function $f$, we denote (as in \cite{qp}) 
by $\undom $ the vector of frequencies defined by 
\be 
     \undom \; := \; \left\{
       \begin{array}{ll}
           \tebom  \in \R^{B}, 
                  & \mbox{if } F_{\teb{0}} = 0 \\
                  & \\
           (\tebom, \; F_{\teb{0}})  \in \R^{B+1}, 
                  & \mbox{if } F_{\teb{0}} \neq 0 ,
         \end{array}
     \right. .
\label{definicaodeundom}
\ee
Since we assume that $ \tebom \in \R_+^B$, the definition above says
that all components of $\undom$ are always non-zero.  Moreover, we
denote
\be
A \; := \;  
     \left\{ 
       \begin{array}{ll}
           B,   & \mbox{if } F_{\teb{0}}   =  0 \\
           & \\
           B+1, & \mbox{if } F_{\teb{0}} \neq 0 
        \end{array}
     \right. .
\label{definicaodeA}
\ee
We denote vectors in $\Z^B$ (or $\R^B $) by $\teb{v}$ and vectors in
$\Z^{A}$ (or $\R^{A} $) by $ \undv$.  The symbol $|\undn|$ denotes the
$l^1(\Z^A)$ norm of a vector $\undn = (n_1, \ldots , n_A)\in \Z^A$: $
|\undn| := |n_1| + \cdots + |n_A|$.  We use the symbol $ \UM$ for the
identity matrix. $\mbox{Mat}(n, \; \C)$ is the set of all $n\times n $
matrices with complex entries. 

We denote by $\lfloor x \rfloor $ the largest integer lower or equal
to $ x\in\R$.

For $ m\in \Z$ we denote by $ \novomod{m}$ the following function:
\be
\novomod{m} \; :=\;  
\left\{ 
  \begin{array}{cl}
                    |m|, & \mbox{for } m \neq
                    0\\ 1 ,& \mbox{for } m =0 
  \end{array}
\right. .
\label{definicaodenovomod}
\ee

For $ m\in \Z$ we denote by $ J_m$ the Bessel function of first
kind and order $ m$.

The symbol $\Box$ denotes end of statement and the symbol
$\rule{2.5mm}{2.5mm}$ denotes end of proof.

\section{Some Previous Results}
\zerarcounters

In \cite{qp} some results could be proven about the nature of some
particular solutions of (\ref{primeiraequacaodeRiccati}) for the case
where $ f$ is a quasi-periodic function subjected to some additional
restrictions.  These results are described in Theorem
\ref{quasiperiodicidadedoscnsEens}.

\begin{theorem}
\label{quasiperiodicidadedoscnsEens}
Let $ f: \R \to \R$ be quasi-periodic with
$$
      f(t) \; = \; \sum_{\teb{n} \in \Z^B} \, F_{\teb{n}} \, 
      e^{i\tebom \cdot \teb{n} t} ,
$$
and such that the sum above contains only a finite number of terms.
Assume that the vector $ \undom$ (defined in (\ref{definicaodeundom}))
satisfies Diophantine conditions, i.e., assume the existence of
constants $\Delta >0 $ and $\sigma >0$ such that, for all $\undn \in
\Z^{A}$, $\undn \neq \und0$,
$$
    |\undn \cdot \undom | \geq \Delta^{-1} |\undn|^{-\sigma} .
$$

\noindent{\bf I.} Assume that $f$ satisfies the condition $M(q^2) \neq 0$.
Then, there exists a formal power series
\be
  g (t) \; = \;
  q(t)\sum_{n=1}^{\infty} \; c_n (t) \eps^n , 
\label{primeiraexpansao}
\ee
representing a particular solution of the generalized Riccati equation
(\ref{primeiraequacaodeRiccati}) such that all coefficients
$c_n$ can be chosen to be quasi-periodic and can be represented as
\be
      c_n(t) \; = \; \sum_{\undm \in \Z^{A}} \, C_{\undm}^{(n)} \, 
      e^{i\undm \cdot \undom t} ,   
\label{definicaodeCnundm}
\ee
where, for the Fourier coefficients $ C_{\undm}^{(n)}$, we have 
$$
    |C_{\undm}^{(n)}| \; \leq \; \calK_n e^{- \chi_0 |\undm|}, 
$$
where $ \chi_0 >0$ is a constant and $\calK_n \geq 0$. 

\noindent{\bf II.} Assume that $f$ satisfies the conditions $M(q^2) = 0$
and $M(\calQ_1) \neq 0$, where $ \calQ_1$ is defined in
(\ref{definicaodecalQ}). Then, there exists a formal power series
\be 
  g (t) \; = \;
  q(t)\sum_{n=1}^{\infty} \; e_n (t) \eps^{2n} , 
\label{segundaexpansao}
\ee
representing a particular solution of the generalized Riccati equation
(\ref{primeiraequacaodeRiccati}) such that all coefficients
$e_n$ can be chosen to be quasi-periodic and can be represented as
\be
      e_n(t) \; = \; \sum_{\undm \in \Z^{A}} \, E_{\undm}^{(n)} \, 
      e^{i\undm \cdot \undom t} ,   
\label{definicaodeEnundm}
\ee
where, for the Fourier coefficients $ E_{\undm}^{(n)}$, we have 
$$
    |E_{\undm}^{(n)}| \; \leq \; \calL_n e^{- \chi_0 |\undm|}, 
$$
where $ \chi_0 >0$ is a constant and $\calL_n \geq 0$. 
$\EndofStatement$

\end{theorem}

There are other conditions beyond {\it I} and {\it II} which could be
considered, but they have not been explored so far. See the discussion
in Section \ref{hierarquias}.

The statements of this last theorem are not sufficient for proving
convergence of the power series expansions in $\eps$ for $ g$.
Unfortunately, as discussed in \cite{qp}, the behavior for large $ n$
of the constants $\calK_n$ and $\calL_n $ is apparently too bad to
guarantee absolute convergence of the formal power series above. 
   
For the restricted case were $f$ is periodic we will prove in the
present paper stronger results (Theorem
\ref{teoremasobredecaimentodosCns} below) than that implied by Theorem
\ref{quasiperiodicidadedoscnsEens}. As we will see, these stronger
results, in contrast, imply convergence of the $ \eps$-power series
for $g$ (Theorem \ref{convergenciadasexpansoesquedaog} below).

Some of the more technical results of \cite{qp} have been obtained
through the analysis of the Fourier coefficients of the functions
$c_n$ and $e_n$ defined in Theorem \ref{quasiperiodicidadedoscnsEens}
above. Specially important for us are the recursion relations found in
\cite{qp} for the Fourier coefficients $ C^{(n)}_{\undm}$ and $
E^{(n)}_{\undm}$ defined in (\ref{definicaodeCnundm}) and
(\ref{definicaodeEnundm}), respectively. Those recursion relations
follow by imposing the generalized Riccati equation
(\ref{primeiraequacaodeRiccati}) to the power expansions
(\ref{primeiraexpansao}) and (\ref{segundaexpansao}).  In Appendix
\ref{ShortDescription} we reproduce some of the main ideas of
\cite{qp} leading to a power series expansion for $ g$ free of secular
terms and leading to the recursion relations below. 
 
It is important for our present purposes to reproduce those recursive
relations here, what we shall do now.

As in \cite{qp}, let us denote by $Q_{\undm}$ the Fourier coefficients
of the function $q$ (defined in (\ref{definicaodeq})) 
\be
       q(t) \; = \; \sum_{m\in \Z} Q_m e^{i m\omega t}
\label{representacaodeFourierdeq}
\ee
and by $Q_{\undm}^{(2)}$ the Fourier coefficients of the function
$q^2$.  For the Fourier coefficients of the functions $ c_n$ we have
found the following relations:
\bear
C_{\undm}^{(1)} & = & \alpha_1 Q_{\undm} ,
\label{CFourier1}
\\
C_{\undm}^{(2)} & = & 
       \sum_{\undn \in \Z^A \atop \undn \neq \und0}
       \frac{\left( \alpha_1^2 Q^{(2)}_{\undn} -
         \overline{Q^{(2)}_{-\undn}} \right)}{\undn \cdot \undom}
     \left[
          Q_{\undm - \undn} - \frac{Q_{\undm} Q_{ -\undn}^{(2)}}{
                             Q_{\und0}^{(2)} }
       \right] ,
\label{CFourier2}
\\
C_{\undm}^{(n)} & = &
       \sum_{\undn_1, \, \undn_2 \in \Z^A \atop \undn_1 + \undn_2 \neq \und0}
       \frac{1}{(\undn_1 + \undn_2) \cdot \undom}
       \left( 
         \sum_{p=1}^{n-1}
         C_{\undn_1}^{(p)}C_{\undn_2}^{(n-p)} 
       \right)
     \left[
          Q_{\undm - (\undn_1 + \undn_2)} - 
          \frac{Q_{\undm} Q_{ -\undn_1 - \undn_2}^{(2)}}{ Q_{\und0}^{(2)} }
       \right]  
\nonumber \\
       & & 
       - \;
       \frac{Q_{\undm}}{2 \alpha_1 Q_{\und0}^{(2)}}
       \sum_{\undn \in \Z^A } \sum_{p=2}^{n-1}
         C_{\undn}^{(p)} C_{-\undn}^{(n+1-p)} , 
         \qquad \qquad \qquad  \mbox{for } n \geq 3.
\label{CFouriern}
\eear

Above $\undm\in\Z^A $, $\displaystyle \alpha_1^2 =
\frac{\overline{M(q^2)}}{M(q^2)}$.  For the Fourier coefficients of
the functions $e_n$ we have found the following relations.
\bear
E_{\undm}^{(1)}
 & = &  
  \sum_{\undn \in \Z^A \atop \undn \neq \und0}
  \frac{ Q_{\undm + \undn} \overline{Q_{ \undn}^{(2)}} }{ \undn \cdot \undom }
  +
  \frac{ Q_{\undm} }{ 2 i M(\calQ_1) }
\sum_{
   \undn_1, \; \undn_2 \in \Z^A \atop \undn_1 \neq \und0 , \; \undn_2
   \neq \und0 
}
\; 
\frac{ 
   Q^{(2)}_{\undn_1 + \undn_2 } \;
   \overline{ Q^{(2)}_{\undn_1 }} \; \overline{ Q^{(2)}_{\undn_2 }} 
     }{ (\undn_1 \cdot \undom ) (\undn_2 \cdot \undom ) }    
\label{EFourier1}
\\
E_{\undm}^{(n)} & = & 
  \sum_{\undn_1 , \, \undn_2 \in \Z^A \atop \undn_1 + \undn_2 \neq  \und0}
\left[
Q_{\undm - \undn_1 - \undn_2}
+
\frac{ Q_{\undm} }{ i M(\calQ_1) }  
   \left(
              Q_{- \undn_1 - \undn_2}^{(2)} \calR
           - \sum_{\undn \in \Z^A \atop \undn \neq \und0}
             \frac{
               Q_{\undn + \undn_1 + \undn_2}^{(2)} 
               \overline{Q_{\undn}^{(2)}} }{
                  \undn \cdot \undom }         
   \right)
\right]
  \frac{\displaystyle
         \sum_{p=1}^{n-1} E_{\undn_1}^{(p)} E_{\undn_2}^{(n-p)} }{(\undn_1 +
          \undn_2)\cdot \undom}
\nonumber
\\
 & & 
    + \frac{ Q_{\undm} }{ 2 i M(\calQ_1) }  
       \sum_{ \undn \in \Z^A } \sum_{p=2}^{n-1}  
       E_{\undn}^{(p)} E_{-\undn}^{(n+1-p)},
       \qquad\qquad\qquad \mbox{ for } n\geq 2.
\label{EFouriern}
\eear
Above $\undm\in\Z^A $, $ \calQ_1$ is defined in
(\ref{definicaodecalQ}) and 
\be
\calR \; := \; 
\frac{1}{2i M(\calQ_1)}
\sum_{
   \undn_1, \; \undn_2 \in \Z^A \atop \undn_1 \neq \und0 , \; \undn_2
   \neq \und0 
}
\; 
\frac{ 
   Q^{(2)}_{\undn_1 + \undn_2 } \;
   \overline{ Q^{(2)}_{\undn_1 }} \; \overline{ Q^{(2)}_{\undn_2 }} 
     }{ (\undn_1 \cdot \undom ) (\undn_2 \cdot \undom ) } .
\ee

The above expressions for the Fourier coefficients are somewhat
complex but two important features can be distinguished. The first is
the inevitable presence of ``small denominators'', represented by the
various factors of the form $(\undn \cdot \undom )^{-1}$ (with $\undn
\neq \und0$) appearing above. The second is the presence of
convolution products (a consequence, lately, of the quadratic
character of the generalized Riccati equation).  The presence of the
later is the additional source of complications mentioned before, for
they also, together with the small denominators, contribute to spoil
the decay of the Fourier coefficients needed to prove convergence of
the $\eps$-expansions.

\section{The Recursive Relations in the Periodic Case}
\zerarcounters

In \cite{qp} the recursion relations presented above have been used to
prove inductively exponential bounds for the Fourier coefficients. As
mentioned before two main difficulties have to be faced in this
enterprise: the presence of ``small denominators'' and of convolution
products in the recursion relations. Both are independently
responsible for reducing the rate of decay of the Fourier coefficients
at each induction step.

Let us consider the origin of the ``small denominators problem'' in
our recursion relations. It comes from the many factors of the form
$(\undn \cdot \undom )^{-1}$ (with $\undn \neq \und0$) appearing in
the recursion relations.  In the case where $f$ is a periodic function
with frequency $\omega$ with $ F_0 \neq 0$, we have $A = 2 $, $\undn =
(n_1, \; n_2) \in \Z^2$ and $\undn \cdot \undom = n_1\omega + n_2F_0$.
On the other hand, in the case where $f$ is a periodic function with
frequency $\omega$ and with $ F_0 = 0$, we have $A = 1 $, $\undn = n \in
\Z $ and $\undn \cdot \undom = n\omega$.  To avoid the
quasi-resonant situation where $n_1\omega + n_2F_0$ is small we will,
as mentioned, consider in this paper the case where $ F_0 = 0$.

For the Fourier coefficients of the functions $c_n$, the recursive
relations become
\bear
C_{m}^{(1)} & = & \alpha_1 Q_{m} ,
\label{PerCFourier1}
\\
C_{m}^{(2)} & = & 
       \sum_{n_1 \in \Z \atop n_1 \neq 0}
       \frac{\left( \alpha_1^2 Q^{(2)}_{n_1} -
         \overline{Q^{(2)}_{-n_1}} \right)}{n_1  \omega}
     \left[
          Q_{m - n_1} - \frac{Q_{m} Q_{ -n_1}^{(2)}}{
                             Q_{0}^{(2)} }
       \right] ,
\label{PerCFourier2}
\\
C_{m}^{(n)} & = &
       \sum_{n_1, \, n_2 \in \Z \atop n_1 + n_2 \neq 0}
       \frac{1}{(n_1 + n_2) \cdot \omega}
       \left( 
         \sum_{p=1}^{n-1}
         C_{n_1}^{(p)}C_{n_2}^{(n-p)} 
       \right)
     \left[
          Q_{m - (n_1 + n_2)} - 
          \frac{Q_{m} Q_{ -n_1 - n_2}^{(2)}}{ Q_{0}^{(2)} }
       \right]  
\nonumber \\
       & & 
       - \;
       \frac{Q_{m}}{2 \alpha_1 Q_{0}^{(2)}}
       \sum_{n_1 \in \Z } \sum_{p=2}^{n-1}
         C_{n_1}^{(p)} C_{-n_1}^{(n+1-p)} , 
       \qquad \qquad \qquad \mbox{for } n \geq 3.
\label{PerCFouriern}
\eear
Above $m\in \Z$. 

For the Fourier coefficients of the functions $e_n$ we have:
\bear
E_{m}^{(1)}
 & = &  
  \sum_{n_1 \in \Z \atop n_1 \neq 0}
  \frac{ Q_{m + n_1} \overline{Q_{ n_1}^{(2)}} }{ n_1  \omega }
  +
  \frac{ Q_{m} }{ 2 i M(\calQ_1) }
\sum_{
   n_1, \; n_2 \in \Z \atop n_1 \neq 0 , \; n_2
   \neq 0 
}
\; 
\frac{ 
   Q^{(2)}_{n_1 + n_2 } \;
   \overline{ Q^{(2)}_{n_1 }} \; \overline{ Q^{(2)}_{n_2 }} 
     }{ (n_1 \omega ) (n_2 \omega ) }    
\label{PerEFourier1}
\\
E_{m}^{(n)} & = & 
  \sum_{n_1 , \, n_2 \in \Z \atop n_1 + n_2 \neq  0}
\left[
Q_{m - n_1 - n_2}
+
\frac{ Q_{m} }{ i M(\calQ_1) }  
   \left(
              Q_{- n_1 - n_2}^{(2)} \calR
           - \sum_{n_3 \in \Z \atop n_3 \neq 0}
             \frac{
               Q_{n_3 + n_1 + n_2}^{(2)} 
               \overline{Q_{n_3}^{(2)}} }{
                  n_3 \omega }         
   \right)
\right]
  \frac{\displaystyle\sum_{p=1}^{n-1}E_{n_1}^{(p)} E_{n_2}^{(n-p)} }{(n_1 +
    n_2) \omega}
\nonumber
\\
 & & 
    + \frac{ Q_{m} }{ 2 i M(\calQ_1) }   \sum_{p=2}^{n-1} 
       \sum_{ n_1 \in \Z } E_{n_1}^{(p)} E_{-n_1}^{(n+1-p)},
       \qquad\qquad\qquad \mbox{ for } n\geq 2.
\label{PerEFouriern}
\eear

It is clear here that no ``small divisors'' appear in this case, since
now $|(\undn \cdot \undom )^{-1}| \geq \omega^{-1}$ for $\undn \neq
\und0$. Hence, the convolution products are the only remaining
factors eventually forcing the reduction of the decay rate of the
Fourier coefficients at the successive induction steps. 

In the Section \ref{InductiveBoundsfortheFourierCoefficients} we will
show how the effect of the convolution products can be taken under
control. The result is expressed in the following three theorems.

\begin{theorem}
  Let $ f: \R \to \R$ be periodic with a finite Fourier decomposition
  as in (\ref{primeiradecomposicaodeFourierdef}) and with $ F_0 =0$.
  
  {\bf Case I.}
  Consider the Fourier coefficients $C_m^{(n)}$ satisfying the
  recursion relations (\ref{PerCFourier1}), (\ref{PerCFourier2}) and
  (\ref{PerCFouriern}). Under the hypothesis that $ M(q^2) \neq 0$
  we have
\be
      |C_m^{(n)}| \; \leq \; K_n \, \frac{e^{-\chi |m|}}{\novomod{m}^2} 
\label{oquequeremosprovar}
\ee
for all $ n\in \N$, and all $ m\in \Z$, where $ \chi >0$ is a constant
and  $ \novomod{m}$ is defined in (\ref{definicaodenovomod}).
Above, the coefficients $K_n $ do not depend on $ m$ and satisfy
the recursion relation 
\be 
   K_n \; = \; \calC_2 \left[ \left( 
         \sum_{p=1}^{n-1} K_p K_{n-p} 
       \right)
+
\left( \sum_{p=2}^{n-1} K_p K_{n+1-p}\right)
\right] ,
\label{relacaorecursivadosKns}
\ee 
with $K_1 = K_2 = \calC_1$, where $\calC_1 $ and $\calC_2 $
are positive constants which can be chosen larger than or equal to $ 1$.

  {\bf Case II.}  
  Consider the Fourier coefficients $E_m^{(n)}$ satisfying the
  recursion relations (\ref{PerEFourier1}) and
  (\ref{PerEFouriern}). Under the hypothesis that $ M(q^2) = 0$ and $
  M(\calQ_1) \neq 0$ we have
\be
      |E_m^{(n)}| \; \leq \; K'_n \, \frac{e^{-\chi |m|}}{\novomod{m}^2} 
\label{oquequeremosprovarE}
\ee
for all $ n\in \N$, and all $ m\in \Z$, where $ \chi >0$ is a constant.
Above, the coefficients $K'_n $ do not depend on $ m$ and satisfy
the recursion relation 
\be 
   K'_n \; = \; \calE_2 \left[ \left( 
         \sum_{p=1}^{n-1} K'_p K'_{n-p} 
       \right)
+
\left( \sum_{p=2}^{n-1} K'_p K'_{n+1-p}\right)
\right] ,
\label{relacaorecursivadosKlinhans}
\ee 
with $K'_1 = K'_2 = \calE_1$, where $\calE_1 $ and $\calE_2 $
are positive constants which can be chosen larger than or equal to $ 1$.
$ \EndofStatement$
\label{teoremasobredecaimentodosCns}
\label{teoremasobredecaimentodosEns}
\end{theorem}

Theorem \ref{teoremasobredecaimentodosCns} will be proven in Section
\ref{InductiveBoundsfortheFourierCoefficients}.  The importance of the
recursive definition of the constants $ K_n$ given in
(\ref{relacaorecursivadosKns}) or (\ref{relacaorecursivadosKlinhans})
is expressed in the following theorem, which says that the constants $
K_n$ grow at most exponentially with $ n$. 

\begin{theorem}
  Let the constants $ K_n$ be defined through the recurrence relations
  (\ref{relacaorecursivadosKns}) or
  (\ref{relacaorecursivadosKlinhans}). Then there exist constants $
  K>0$ and $ K_0 >0$ (depending eventually on $ f$) such that $K_n
  \leq K_0 K^n$ for all $n\in \N $.  $ \EndofStatement$
\label{ProposicaotipoCatalan}
\end{theorem}

The proof of Theorem \ref{ProposicaotipoCatalan} is found in Appendix
\ref{ProvadaProposicaotipoCatalan} and makes interesting use of
properties of the Catalan sequence.  Theorems
\ref{teoremasobredecaimentodosCns} and \ref{ProposicaotipoCatalan}
have the following immediate corollary:

\begin{theorem}
  The power series expansions in (\ref{primeiraexpansao}) and
  (\ref{segundaexpansao}) are absolutely convergent for all $\eps\in\C
  $ with $ |\eps |< K$ for all $ t\in\R$ and, hence,
  (\ref{primeiraexpansao}) and (\ref{segundaexpansao}) define
  particular solutions of the generalized Riccati equation
  (\ref{primeiraequacaodeRiccati}) in cases I and II, respectively, of
  Theorem \ref{teoremasobredecaimentodosCns}.  The function $ g$ can
  be expressed in terms of an absolutely and uniformly converging
  Fourier series whose coefficients can be expressed in terms of
  absolutely converging power series in $ \eps$ for all $\eps\in\C $
  with $ |\eps |< K$.  $ \EndofStatement$
\label{convergenciadasexpansoesquedaog}
\end{theorem} 

\noindent {\bf Proof of Theorem \ref{convergenciadasexpansoesquedaog}}
We prove the statement for case I. Case II is analogous. 
The first step is to determine the Fourier expansion
of the function $ g$, as given in (\ref{expansaodeg}), and to study
some of their properties. One clearly has
\be
     g(t) \; = \; \sum_{m\in \Z}G_m e^{i m \omega t}, 
\ee
with
\be
       G_m \; := \; \sum_{n=1}^{\infty}\eps^n  G_m^{(n)},
\ee
where
\be
     G_m^{(n)} \; := \; \sum_{l \in \Z} Q_{m-l} C^{(n)}_{l} .
\label{REvbwpiebpwrert}
\ee

Now and in future proofs we will make use of the following important lemma,
whose proof is given in Appendix \ref{apendice2}.

\begin{lemma}
For $ \chi > 0$ and $ m\in\Z$ define
\be
   \calB (m) \equiv \calB(m, \, \chi) \; := \;  \sum_{n \in \Z } 
\frac{e^{-\chi(|m-n|+|n|)}}{\novomod{m-n}^2\; \novomod{n}^2  } .
\label{definicaodecalB}
\ee
Then one has
\be
   \calB (m) \; \leq \; B_0 \frac{e^{-\chi|m|}}{\novomod{m}^2}
\ee
for some constant $ B_0 \equiv B_0(\chi) >0$ and for all $ m\in \Z$.  
$\EndofStatement$
\label{lemaauxilialsobredecaimentodasconvolucoes}
\end{lemma}

We have the following proposition:

\begin{proposition}
For all $ \chi>0 $
there exists a constant $ \calC_g \equiv \calC_g (\chi)> 0$ such that 
\be
       |G_m^{(n)}| \; \leq \; \calC_g K_n \frac{e^{-\chi |m|}}{\novomod{m}^2}
\label{oidfuvgnpsDsrf}
\ee
for all $m\in \Z$ and all $ n\in \N$. Consequently, for $|\eps| < K$
one has
\be
       |G_m| \; \leq \; \calC'_g \frac{e^{-\chi |m|}}{\novomod{m}^2}
\ee
for some constant $\calC'_g (\chi, \eps) >0$ and for all  $m\in \Z$. 
$\EndofStatement$
\label{propdodecaimentodosGs}
\end{proposition}

\noindent {\bf Proof of Proposition \ref{propdodecaimentodosGs}.} 
Inserting (\ref{oquequeremosprovar}) and (\ref{RR1}) into
(\ref{REvbwpiebpwrert}) we have, for any $ \chi >0$
\be
     \left| G_m^{(n)}\right| \; \leq \; K_n\calQ \; \calB(m, \, \chi) ,
\ee
where $\calB(m, \, \chi) $ is defined in (\ref{definicaodecalB}).
Relation (\ref{oidfuvgnpsDsrf}) follows now from Lemma
\ref{lemaauxilialsobredecaimentodasconvolucoes}. $\EndofProof$

From this the rest of the proof of Theorem
\ref{convergenciadasexpansoesquedaog} follows immediately. $ \EndofProof$

The solutions for the generalized Riccati equation
(\ref{primeiraequacaodeRiccati}) mentioned in Theorem
\ref{convergenciadasexpansoesquedaog} are, through (\ref{definicaodeU}),
the main ingredient for the solution of the Schr\"odinger equation
(\ref{schroedingerPhi}).  This will be further discussed in Section
\ref{TheWaveFunctions}. Now we have to prove Theorem
\ref{teoremasobredecaimentodosCns}.

\section{Inductive Bounds for the Fourier Coefficients}
\zerarcounters
\label{InductiveBoundsfortheFourierCoefficients}

In this section we will prove Theorem
\ref{teoremasobredecaimentodosCns} in cases I and II.  We will make
use of the following proposition on the decay of the Fourier
coefficients $ Q_m$ and $Q_m^{(2)} $ of the functions $ q$ and $ q^2$,
respectively. The proof of this proposition appears in Appendix
\ref{apendice1}.

\begin{proposition}
  Let $ f :\R \to\R$ be periodic and be represented by a finite
  Fourier series as in (\ref{primeiradecomposicaodeFourierdef}).
  Then, for any constant $ \chi >0$ there is a positive constant $
  \calQ \equiv \calQ(\chi)$ such that
\be
      |Q_m| \; \leq \; \calQ \frac{e^{-\chi |m|}}{\novomod{m}^2} 
\label{RR1}
\ee
and
\be
      |Q_m^{(2)}| \; \leq \; \calQ \frac{e^{-\chi |m|}}{\novomod{m}^2} 
\label{RR2}
\ee
for all $ m\in \Z$, where
$ \novomod{m}$ is defined in (\ref{definicaodenovomod}).
$\EndofStatement$
\label{proposicaosobreodecaimentodeQm}
\end{proposition}

\subsection{Case I}

In this section we will prove Theorem
\ref{teoremasobredecaimentodosCns} in case I. Making use of
Proposition \ref{proposicaosobreodecaimentodeQm} and of relations
(\ref{PerCFourier1}), (\ref{PerCFourier2}) and (\ref{PerCFouriern}) we
easily derive the following estimates:
\bear
|C_{m}^{(1)}| & \leq & \calQ \frac{e^{-\chi|m|}}{\novomod{m}^2} ,
\label{Iq1}
\\
|C_{m}^{(2)}| & \leq & 
       2\omega^{-1}\calQ\sum_{n_1 \in \Z }
      \frac{e^{-\chi|n_1|}}{\novomod{n_1}^2}
     \left[
          \frac{e^{-\chi|m - n_1|}}{\novomod{m-n_1}^2} + 
           \frac{\calQ}{|Q_{0}^{(2)}|} \frac{e^{-\chi(|m|+|n_1|)}}{
                             \novomod{m}^2\; \novomod{n_1}^2 }
       \right] ,
\label{Iq2}
\\
|C_{m}^{(n)}| & \leq &
       \omega^{-1}\calQ \sum_{n_1, \, n_2 \in \Z }
       \left( 
         \sum_{p=1}^{n-1}
         |C_{n_1}^{(p)}|\, |C_{n_2}^{(n-p)}| 
       \right)
     \left[
          \frac{e^{-\chi|m - (n_1 + n_2)|}}{\novomod{m - (n_1 + n_2)}^2} + 
          \frac{\calQ}{|Q_{0}^{(2)}|}
          \frac{e^{-\chi(|m|+|n_1+n_2|)}}{\novomod{m}^2\; 
                                                 \novomod{n_1+n_2}^2  }
       \right]  
\nonumber \\
       & & 
       + \;
       \frac{\calQ}{2 |Q_{0}^{(2)}|}
       \frac{e^{-\chi|m|}}{\novomod{m}^2}
       \sum_{n_1 \in \Z } \sum_{p=2}^{n-1}
         |C_{n_1}^{(p)}|\, | C_{-n_1}^{(n+1-p)}| , 
       \qquad \qquad \qquad \mbox{for } n \geq 3.
\label{Iqn}
\eear

It follows from (\ref{Iq2}), from the definition of $ \calB(m)$ in
(\ref{definicaodecalB}) and from Lemma
\ref{lemaauxilialsobredecaimentodasconvolucoes} that

\be
|C_{m}^{(2)}|  \leq 
       2\omega^{-1}\calQ \left( \calB(m) +  
\frac{\calQ}{|Q_{0}^{(2)}|} \frac{e^{-\chi|m|}}{
                             \novomod{m}^2 }
\sum_{n_1 \in \Z}\frac{e^{-2\chi|n_1|}}{
                             \novomod{n_1}^4 }\right)
\leq
K_2\frac{e^{-\chi|m|}}{\novomod{m}^2 }
\label{duhpwer}
\ee
for some convenient choice of the constant $K_2$.

Now, we will use an induction argument to establish
(\ref{oquequeremosprovar}) for all $ n \geq 3$. Let us assume that, for
a given $ n\in \N$, $n \geq 3 $, one has
\be
      |C_m^{(p)}| \; \leq \; K_p \, \frac{e^{-\chi |m|}}{\novomod{m}^2} ,
      \qquad \forall m\in \Z ,
\label{ekjrhpvevsege}
\ee
for all $p$ such that $1 \leq p \leq n-1$, for some convenient
constants $K_p$. We will establish that this implies the same sort of
bound for $p=n$. Notice, by taking $ K_1 \geq \calQ$, that relation
(\ref{Iq1}) guarantees (\ref{ekjrhpvevsege}) for $p=1$ and that
relation (\ref{duhpwer}) guarantees the case $ p=2$.

From (\ref{Iqn}) and from the induction hypothesis,
\bear
|C_{m}^{(n)}| & \leq &
       \omega^{-1}\calQ 
       \left( 
         \sum_{p=1}^{n-1} K_p K_{n-p} 
       \right)
     \Bigg[
      \sum_{n_1, \, n_2 \in \Z }
          \frac{e^{-\chi(|m - (n_1 + n_2)|+|n_1|+|n_2|)}}{
          \novomod{m - (n_1 + n_2)}^2 \; \novomod{n_1}^2\; \novomod{n_2}^2} 
  \nonumber \\ 
          & + & 
          \frac{\calQ}{|Q_{0}^{(2)}|}
          \frac{e^{-\chi|m|}}{\novomod{m}^2}
          \sum_{n_1, \, n_2 \in \Z }
          \frac{e^{-\chi(|n_1+n_2|+|n_1|+|n_2|)}}{
          \novomod{n_1+n_2}^2\; \novomod{n_1}^2\; \novomod{n_2}^2  }
       \Bigg]  
  \nonumber \\
       & + & 
       \frac{\calQ}{2 |Q_{0}^{(2)}|}
       \frac{e^{-\chi|m|}}{\novomod{m}^2}
       \left( \sum_{p=2}^{n-1} K_p K_{n+1-p}\right)
       \sum_{n_1 \in \Z } \frac{e^{-2\chi|n_1|}}{\novomod{n_1}^4 } .
\eear
Now,
$$
          \sum_{n_1, \, n_2 \in \Z }
          \frac{e^{-\chi(|n_1+n_2|+|n_1|+|n_2|)}}{
          \novomod{n_1+n_2}^2\; \novomod{n_1}^2\; \novomod{n_2}^2  }
\qquad \mbox{ and } \qquad 
\sum_{n_1 \in \Z } \frac{e^{-2\chi|n_1|}}{\novomod{n_1}^4 }
$$
are just finite constants and
\bear
      \sum_{n_1, \, n_2 \in \Z }
          \frac{e^{-\chi(|m - (n_1 + n_2)|+|n_1|+|n_2|)}}{
          \novomod{m - (n_1 + n_2)}^2 \; \novomod{n_1}^2\; \novomod{n_2}^2}
& = &
\sum_{n_1  \in \Z }
          \frac{e^{-\chi|n_1|}}{\novomod{n_1}^2}
\sum_{n_2 \in \Z }         
         \frac{e^{-\chi(|(m - n_1) - n_2)|+|n_2|)}}{
          \novomod{(m - n_1) - n_2)}^2 \; \novomod{n_2}^2}
\nonumber \\
& = &
\sum_{n_1  \in \Z }
          \frac{e^{-\chi|n_1|}}{\novomod{n_1}^2}\calB(m-n_1)
\nonumber \\
& \leq &
B_0 \sum_{n_1  \in \Z }
          \frac{e^{-\chi(|n_1|+|m-n_1|)}}{\novomod{n_1}^2\; \novomod{m-n_1}^2}
\nonumber \\
& = &
B_0 \calB(m)
\nonumber \\
& \leq &
(B_0)^2 \frac{e^{-\chi|m|}}{\novomod{m}^2} ,
\eear
where we again used Lemma
\ref{lemaauxilialsobredecaimentodasconvolucoes}. 

Therefore, we conclude
\be
|C_{m}^{(n)}|  \; \leq \;  
\left[
\calC_a \left( 
         \sum_{p=1}^{n-1} K_p K_{n-p} 
       \right)
+
\calC_b
\left( \sum_{p=2}^{n-1} K_p K_{n+1-p}\right)
\right]\;
\frac{e^{-\chi|m|}}{\novomod{m}^2},
\ee
for two positive constants $\calC_a$ and $\calC_b$. Taking $
\calC_2 := \max \{ \calC_a, \; \calC_b, \; 1\}$ relation
(\ref{relacaorecursivadosKns}) is proven with $\calC_2 \geq 1$.  

Notice that, without loss, we are allowed to choose $ K_1 = K_2 \geq 1$
by choosing both equal to $\max \{K_1, \; K_2, \; 1\}$. 
$\EndofProof$

\subsection{Case II}

In this section we will prove Theorem
\ref{teoremasobredecaimentodosCns} in case II.  From
(\ref{PerEFourier1}) and (\ref{PerEFouriern}), from Proposition
\ref{proposicaosobreodecaimentodeQm} and from the assumption
(\ref{oquequeremosprovarE}) we have
\bear
\left| E_{m}^{(1)} \right|
 & \leq &  
  \frac{\calQ^2}{\omega}\sum_{n_1 \in \Z}
\frac{ e^{-\chi(|m+n_1|+|n_1|)} }{\novomod{m+n_1}^2 \, \novomod{n_1}^2 } 
\nonumber
\\ & + &
  \frac{\calQ^4  e^{-\chi|m|} }{ 2\novomod{m}^2 \omega^2 |M(\calQ_1)| }
\sum_{
   n_1, \; n_2 \in \Z }
\; 
\frac{ 
e^{-\chi(|n_1+n2|+|n_1|+|n_2|) }
}{
\novomod{n_1 + n_2}^2 \, \novomod{n_1}^2 \, \novomod{n_2}^2 
} \; \; \; \; ,
\\
E_{m}^{(n)} & = & 
 \frac{1}{\omega} \sum_{n_1 , \, n_2 \in \Z }
\Biggl[
\calQ\frac{e^{-\chi(|m - n_1 - n_2|+|n_1|+|n_2|)}
}{
   \novomod{m - n_1 - n_2}^2\, \novomod{n_1}^2\, \novomod{n_2}^2
}
\nonumber
\\ & &
+
\frac{\calQ^2 e^{-\chi|m|}  }{ |M(\calQ_1)|\novomod{m}^2 }  
   \Biggl(
     \frac{e^{-\chi(| n_1 + n_2|+|n_1|+|n_2|)}|\calR|
    }{
     \novomod{ n_1 + n_2}^2\, \novomod{n_1}^2\, \novomod{n_2}^2
    } 
\nonumber
\\ & &
 + \frac{\calQ}{\omega}\sum_{n_3 \in \Z }
   \frac{e^{-\chi(|n_1+n_2+n_3|+|n_1|+|n_2|+|n_3|)}
   }{
   \novomod{n_1+n_2+n_3}^2\,\novomod{n_1}^2\, \novomod{n_2}^2\,
   \novomod{n_3}^2
   } 
   \Biggr)
\Biggr]
  \left(\sum_{p=1}^{n-1}\; K'_p K'_{n-p}\right)
\nonumber
\\
 & & 
 +  \frac{\calQ e^{-\chi|m|} }{ 2 |M(\calQ_1)|\novomod{m}^2 }   
    \left(\sum_{ n_1 \in \Z }\frac{e^{-2\chi|n_1|}}{\novomod{n_1}^4}\right) \;
    \left(\sum_{p=2}^{n-1} \; K'_p K'_{n+1-p}\right),
       \qquad\mbox{ for } n\geq 2.
\eear

Sums like
$$
\sum_{
   n_1, \; n_2 \in \Z }
\; 
\frac{ 
e^{-\chi(|n_1+n2|+|n_1|+|n_2|) }
}{
\novomod{n_1 + n_2}^2 \, \novomod{n_1}^2 \, \novomod{n_2}^2 
}
\; \mbox{ and } \;
\sum_{
   n_1, \; n_2, \; n_3 \in \Z }
   \frac{e^{-\chi(|n_1+n_2+n_3|+|n_1|+|n_2|+|n_3|)}
   }{
   \novomod{n_1+n_2+n_3}^2\,\novomod{n_1}^2\, \novomod{n_2}^2\,
   \novomod{n_3}^2
   } 
$$
are just finite constants. By applying Lemma
\ref{lemaauxilialsobredecaimentodasconvolucoes} we get
\bear
|E_m^{(1)}| & \leq & \calE_a \frac{e^{-\chi|m|}}{\novomod{m}^2}  
\\
|E_m^{(n)}| & \leq & \frac{e^{-\chi|m|}}{\novomod{m}^2}
\left[
\calE_b  \left(\sum_{p=1}^{n-1}\; K'_p K'_{n-p}\right)
+
\calE_c \left(\sum_{p=2}^{n-1} \; K'_p K'_{n+1-p}\right)
\right]  ,
       \qquad \mbox{ for } n\geq 2,
\eear
where $ \calE_a$, $\calE_b$ and $\calE_c $ are constants.  The rest of
the proof follows the same steps of the proof of Theorem
\ref{teoremasobredecaimentodosCns} in case I. $\EndofProof$

\section{The Fourier Expansion for the Wave Function}
\zerarcounters
\label{TheWaveFunctions}

Now we return to the discussion of the solution (\ref{definicaodeU}) of the
Schr\"odinger equation (\ref{schroedingerPhi}). Our intention is to
find the Fourier expansion of the wave function $ \Phi(t)$. 

\subsection{The Floquet Form of the Wave Function. The Fourier
  Decomposition and the Secular Frequency}

As explained in \cite{qp} and in Section \ref{Introduction}, the
components $ \phi_\pm$ of the wave function $ \Phi(t)$ are solutions
of Hill's equation (\ref{Hillparapm}). For periodic $ f$ the classical
theorem of Floquet (see e.g. \cite{Heuser3} and \cite{Hochstadt})
claims that there are particular solutions of equations like
(\ref{Hillparapm}) with the general form
$
   e^{i\Omega t}u(t) 
$, 
where $u(t)$ is periodic with the same period of $ f$. In order
to preserve unitarity we must have $ \Omega \in \R$. This form of the
particular solutions is called the ``Floquet form'' and the frequencies
$\Omega$ are called ``secular frequencies''. 

In this section we will recover the Floquet form of the wave function
in terms of Fourier expansions and we will find out expansions for the
secular frequencies as converging power series expansions in $ \eps$.

According to the solution expressed in relation
(\ref{solucaodaequacaodeSchroedinger}) and (\ref{definicaodeU}), we have
first to find out the Fourier expansion for the functions $ R$ and $
S$ defined in (\ref{definicaodeR}) and (\ref{definicaodeS}),
respectively.

We begin with the function $ R$. The Fourier expansion of the function
$ f+g$ is
\be
  f(t)+g(t)  \; = \;  \Omega + \sum_{n\in \Z \atop n\neq 0} 
                 (F_n + G_n(\eps)) \, e^{in\omega t}, 
\ee
where
\be
        \Omega \; \equiv \; \Omega(\eps) \; := \;  G_0(\eps).  
\ee
One has,
\be
    R(t) \; = \;  e^{-i\gamma_f(\eps)}\;e^{-i \Omega t}\;
    \exp \left(
                -\sum_{n\in \Z } H_n e^{in\omega t}
         \right)
\label{isuybvgpweruewp}
\ee
with
\be
H_n \; \equiv \; H_n(\eps)\; := \; \left\{ \begin{array}{cl}
          \displaystyle \frac{F_n + G_n(\eps)}{n\omega} , 
                                       & \mbox{for } n\neq 0 \\
                                       & \\
                             0,        & \mbox{for } n=0
               \end{array}
         \right. ,
\ee 
and
\be
     \gamma_f(\eps) \; := \;  i \sum_{m\in \Z } \, H_m .
\ee
Notice that $\gamma_f(0) = \gamma_f $, where $ \gamma_f$ is defined in 
(\ref{definicaodegamaf}). 

Since we are assuming that there are only finitely many non-vanishing
coefficients $F_n$, we have the following proposition as an obvious
corollary of Proposition \ref{propdodecaimentodosGs}:

\begin{proposition}
For all $ \chi>0 $ and $|\eps| $ small enough, there exists a constant 
$ \calC_H \equiv \calC_H(\chi, \, \eps)> 0$ such that 
\be 
    |H_m| \; \leq \; \calC_H \frac{e^{-\chi |m|}}{\novomod{m}^2} 
\ee 
for all $m\in \Z$.  $\EndofStatement$
\label{propdodecaimentodosHs}
\end{proposition}

Writing now the Fourier expansion of $ R(t)$ in the form
\be
       R(t) \; = \; e^{-i \Omega t}\sum_{n\in \Z } R_n e^{in\omega t}
\ee
we find from (\ref{isuybvgpweruewp})
\be
    R_n \equiv  R_n(\eps) = 
\left\{
\begin{array}{ll}
 \displaystyle
 e^{-i\gamma_f(\eps)}
 \left( 
       1 +  \sum_{p=1}^{\infty}\frac{(-1)^{p+1}}{(p+1)!} 
            \sum_{n_1 , \ldots ,\,  n_p \in \Z}
                              H_{n_1} \cdots H_{n_p} H_{ -N_p}
 \right), & \mbox{for } n = 0, \\
 & \\
 \displaystyle
 e^{-i\gamma_f(\eps)}
 \left(
        -H_n +  \sum_{p=1}^{\infty}\frac{(-1)^{p+1}}{(p+1)!} 
                \sum_{n_1 , \ldots , \, n_p \in \Z}
                              H_{n_1} \cdots H_{n_p} H_{ n - N_p}
 \right), & \mbox{for } n \neq 0 .
\end{array}
\right.
\label{definicaodosRns}
\ee
with
\be
       N_p \; := \; \sum_{a=1}^{p}n_a ,
\ee
for $ p\geq 1$. 

In order to compute the Fourier expansion of $S$ we have to compute
first the Fourier expansion of $R^{-2}$. This is now an easy task,
since the replacement $R(t) \to R(t)^{-2} $ corresponds to the replacement
$ (f+g) \to -2(f+g)$ and, hence, to $ H_n \to -2H_n$. We get
\be
    R(t)^{-2} \; = \; e^{2i\Omega t}\sum_{n\in \Z }R_n^{(-2)}e^{in\omega t},
\ee
with
\be
R_n^{(-2)} \; \equiv \; R_n^{(-2)}(\eps) \; := \; 
\left\{
\begin{array}{ll}
 \displaystyle
 e^{2i\gamma_f(\eps)}
 \left( 
       1 +  \sum_{p=1}^{\infty}\frac{2^{p+1}}{(p+1)!} 
            \sum_{n_1 , \ldots , \, n_p \in \Z}
                              H_{n_1} \cdots H_{n_p} 
                              H_{ -N_p}
 \right), & \mbox{for } n=0 , \\
 & \\
 \displaystyle
 e^{2i\gamma_f(\eps)}
 \left(
        2H_n +  \sum_{p=1}^{\infty}\frac{2^{p+1}}{(p+1)!} 
                 \sum_{n_1 , \ldots , \, n_p \in \Z}
                              H_{n_1} \cdots H_{n_p}
                              H_{ n - N_p}
 \right), & \mbox{for } n\neq0 .
\end{array}
\right. 
\ee

The following proposition will be used below.

\begin{proposition}
For all $ \chi>0 $ and $|\eps| $ small enough, there exist constants 
$ \calC_R \equiv \calC_R (\chi, \, \eps)> 0$ and 
$ \calC_{R^{(-2)}} \equiv \calC_{R^{(-2)}}(\chi, \, \eps)> 0$
such that 
\be 
  |R_m| \; \leq \; \calC_R  \frac{e^{-\chi |m|}}{\novomod{m}^2} 
\ee 
\be 
  |R^{(-2)}_m| \; \leq \; \calC_{R^{(-2)}} \frac{e^{-\chi |m|}}{\novomod{m}^2} 
\ee 
for all $m\in \Z$.  $\EndofStatement$
\label{propdodecaimentodosRs}
\end{proposition}

\noindent {\bf Proof of Proposition \ref{propdodecaimentodosRs}.} 
Using Proposition \ref{propdodecaimentodosHs} we have, for any 
$p\geq 1$,
\be
\left|
 \sum_{n_1 , \ldots , \, n_p \in \Z}
                              H_{n_1} \cdots H_{n_p} \, H_{n -N_p}
\right|
\leq
(\calC_H)^{p+1}  \sum_{n_1 , \ldots , \, n_p \in \Z}
\frac{
      \exp\left( -\chi(|n_1|+\cdots+|n_p| +|n - n_1 - \cdots -n_p| )\right)
}{
  \left(
        \novomod{n_1} \cdots \novomod{n_p}\, \novomod{n - n_1 - \cdots -n_p}
  \right)^2
} .
\label{sdifvnIUsoiun}
\ee
Making repeated use of Lemma
\ref{lemaauxilialsobredecaimentodasconvolucoes} on the right hand side
of (\ref{sdifvnIUsoiun}) we get 
\be
\left|
 \sum_{n_1 , \ldots , \, n_p \in \Z}
                              H_{n_1} \cdots H_{n_p} \, H_{n -N_p}
\right|
\; \leq \; 
\frac{(\calC_H B_0)^{p+1}}{B_0} \frac{e^{-\chi|n|}}{\novomod{n}^2} .
\ee
Inserting this into (\ref{definicaodosRns}) gives (since $B_0 >1$)
\be
 |R_n| \;\;\; \leq \;\;\; 
 \left( 
      \frac{e^{\left|\mbox{\scriptsize Im}(\gamma_f (\eps))\right|+\calC_H
          B_0}}{B_0}
 \right)\; \;
     \frac{e^{-\chi|n|}}{\novomod{n}^2}
\ee
for all $ n\in \Z$, as desired. The proof for $R_n^{(-2)}$ is
analogous.  $\EndofProof$

Assuming for a while
\be 
n\omega +
2\Omega \neq 0 \qquad \mbox{ for all } n\in \Z ,
\label{condicaodenaocancelamentodosdenominadores} 
\ee 
we have\footnote{For the case $ n=0$,
  (\ref{condicaodenaocancelamentodosdenominadores}) says that $ \Omega
  \neq 0$. This must hold except for $ \eps=0$ when $ \Omega =0$. } 
\be
     S(t) \; = \; \sigma_0 + e^{2i\Omega t}\sum_{n\in\Z} S_n
     e^{in\omega t}
\label{Sdet}
\ee
with
\be
     S_n \; := \; -i\frac{R_n^{(-2)}}{ n\omega + 2\Omega}
\qquad \mbox{ and } \qquad
     \sigma_0 \; := \; - \sum_{n\in\Z}\,  S_n .
\ee
Assumption (\ref{condicaodenaocancelamentodosdenominadores} ) is
actually a consequence of unitarity, as will be discussed in Section
\ref{unitariedade}.

The following proposition is an elementary corollary of Proposition 
\ref{propdodecaimentodosRs}:

\begin{proposition}
For all $ \chi>0 $ and $|\eps| $ small enough, there exists a constant 
$ \calC_S \equiv \calC_S (\chi, \, \eps)> 0$ 
such that 
\be 
  |S_m| \; \leq \; \calC_S  \frac{e^{-\chi |m|}}{\novomod{m}^2} 
\ee 
for all $m\in \Z$.  $\EndofStatement$
\label{propdodecaimentodosSs}
\end{proposition}

Writing 
\be
U(t) \; = \; 
\left(
 \begin{array}{cc}
    U_{11}(t) & U_{12}(t) \\  & \\ U_{21}(t) & U_{22}(t) 
   \end{array}
\right)
\; = \;
\left(
 \begin{array}{cc}
    U_{11}(t) & U_{12}(t) \\ & \\ -\overline{U_{12}(t)} & \overline{U_{11}(t)} 
   \end{array}
\right) ,
\label{duynhgsd}
\ee
we have for $ U_{11}$ and $ U_{12}$:
\bear
U_{11}(t) & = & e^{-i\Omega t}\, u_{11}^-(t) + e^{i\Omega t}\, u_{11}^+ (t)
\label{U11dffLJ}
\\
U_{12}(t) & = & e^{-i\Omega t}\, u_{12}^- (t) + e^{i\Omega t}\, u_{12}^+ (t)
\label{U12dffLJ}
\eear
with 
\be
\begin{array}{lclcclcl}
u_{11}^- (t)  & := &  (1+i g(0)\sigma_0) \, r(t),
 & & &
u_{11}^+ (t)  & := & i g(0)\, v(t),
\\ & & \\
u_{12}^- (t)  & := & -i\eps\sigma_0 \, r(t),
 & & &
u_{12}^+ (t)  & := & -i\eps \, v(t) ,
\end{array}
\ee
for 
\be
    r(t) \; := \; \sum_{n\in\Z} R_n \, e^{in\omega t}
\qquad
\mbox{ and }
\qquad
    v(t) \; := \; \sum_{n\in\Z}
                 \, V_n \, e^{in\omega t},
\ee
with
\be
       V_n \; := \; \sum_{m\in\Z}S_{n-m}R_m .
\ee

This provides the desired Floquet form for the components of the wave
function $ \Phi(t)$. We notice from the expressions above that the
secular frequencies are $ \pm \Omega$. For $ \Omega$ we have the
$\eps$-expansion
\be
    \Omega =  \sum_{n=1}^{\infty} \eps^n G^{(n)}_{0} ,
\label{expansaodeOmega}
\ee
and for $ g(0)$,
\be
       g(0) \; = \; \sum_{m\in \Z} G_m  \; = \; 
                    \sum_{n=1}^{\infty}\eps^n \sum_{m\in \Z} G_m^{(n)} .
\ee
Both converge absolutely for $ |\eps| < K$, where $ K$ is mentioned
in Theorem \ref{ProposicaotipoCatalan}.

As before, we have the following corollary of Propositions 
\ref{propdodecaimentodosRs}, \ref{propdodecaimentodosSs} and Lemma
\ref{lemaauxilialsobredecaimentodasconvolucoes}:

\begin{proposition}
For all $ \chi>0 $ and $|\eps| $ small enough, there exists a constant 
$ \calC_V \equiv \calC_V (\chi, \, \eps)> 0$ 
such that 
\be 
  |V_m| \; \leq \; \calC_V  \frac{e^{-\chi |m|}}{\novomod{m}^2} 
\ee 
for all $m\in \Z$.  $\EndofStatement$
\label{propdodecaimentodosVs}
\end{proposition}

This last proposition closed the proof of Theorem
\ref{Teoremaprincipalsobreopropagador}.

\subsection{Remarks on the Unitarity of the Propagator}
\label{unitariedade}

The unitarity of the propagator $U(t) $ means $ U(t)^*U(t) =
\UM$. After (\ref{duynhgsd}), this means
\be
     |U_{11}(t)|^2 +|U_{12}(t)|^2 \; = \; 1 .
\label{condicaoreduzidadeunitariedade}
\ee

Looking at relations (\ref{U11dffLJ}) and (\ref{U12dffLJ}) two
conclusions can be drawn from (\ref{condicaoreduzidadeunitariedade}).
The first is the following proposition:
\begin{proposition}
  For $ \eps \in \R$ and under the hypothesis leading to
  (\ref{U11dffLJ}) and (\ref{U12dffLJ}) one has $\Omega \in \R $. $
  \EndofStatement$
\label{Omegaehreal}
\end{proposition}

The proof follows from the obvious observation that
(\ref{condicaoreduzidadeunitariedade}) would be violated for $|t|$
large enough if $ \Omega$ had a non-vanishing imaginary part.
Unfortunately a proof of this fact using directly the $
\eps$-expansion of $ \Omega$ (\ref{expansaodeOmega}) is difficult and
has not been found yet.

The second conclusion is that
(\ref{condicaodenaocancelamentodosdenominadores}) indeed holds. For,
without this assumption there would be a term linear in $ t$ in
(\ref{Sdet}), violating (\ref{condicaoreduzidadeunitariedade}) for
large $ |t|$.

As in the case of Proposition \ref{Omegaehreal}, no direct proof of
this fact out of the $ \eps$-expansion for $ \Omega$
(\ref{expansaodeOmega}) has been found yet. The proof will probably
follow the idea that $|\Omega|$ is always smaller than $ 2\omega$
because $\Omega $ is of order $ \eps$ and $|\eps |$ has to be chosen
small in order to provide convergence for the expansions. Analogously
$ \Omega \neq 0$ because $\Omega$ is analytic in $\eps$ and, hence,
has isolated zeros. If the analyticity domain must be small enough no
zeros occur, except at $ \eps =0$.

\section{Discussion on the Classes of Solutions}
\zerarcounters
\label{hierarquias}

Let us now discuss some aspects of conditions I and II of Theorem
\ref{Teoremaprincipalsobreopropagador}. It is important to stress that
these conditions are restrictions on the function $ f$ and not on the
parameter $\eps$.

As in (\ref{efugbyhwvpe}), let us write the Fourier decomposition of $
f$ as
\be 
    f(t) = \sum_{a=1}^{2J} f_a e^{i n_a \omega t}, 
\ee 
with $n_a = - n_{2J - a+1}$ and $\overline{f_a} = f_{2J -a +1}$ for
all $ a$ with $1\leq a \leq J$. 
Comparing with (\ref{primeiradecomposicaodeFourierdef}) one has $f_a
\equiv F_{n_a}$, $1\leq a \leq J$.

Hence, for $ F_0=0$ and for fixed $ J$ and $ \omega$, there are $ J$
independent complex coefficients $f_a$ and we can identify the
parameter space $ \R^{2J}$ with the set ${\mathfrak F}_{J, \, \omega}
$ of all possible functions $f$ with a given $ J$ and $ \omega$.

Condition $ M(q^2)=0$ determines a $(2J-1)$ or $(2J-2)$-dimensional
subset of $ {\mathfrak F}_{J, \, \omega}$ and there condition II
applies.  It is also on this subset that the more restrictive
condition $ M(q^2) = M(\calQ_1)=0$ should hold, restricting the
parameter space of $ f$ to a $(2J-2)$, $(2J-3) $ or
$(2J-4)$-dimensional subset. Hence, successive conditions like I and
II would eventually exhaust completely the parameter space $
{\mathfrak F}_{J, \, \omega}$.

Conditions beyond I and II have not been yet analysed and many
questions concerning the classes of solutions are still open. For
instance, will further conditions like I and II really exhaust the
parameter space of the functions $f $? Will the subtraction method of
\cite{qp} and the convergence proofs of the present paper also work
under these further conditions?  What are the physically qualitative
distinctions between the classes?  Are these classes of solutions in
some sense analytic continuations of each other?

A distinction between class I and II may be pointed with the
observation that in class I we have power expansions in $\eps$ while
in II we have power expansions in $\eps^2 $. Compare relations
(\ref{primeiraexpansao}) and (\ref{segundaexpansao}) of Theorem
\ref{quasiperiodicidadedoscnsEens}.

\subsection{An Explicit Example}

To illustrate these ideas and point to some problems let us consider
the important example where $ f$ is given by
\be
      f(t) \; = \; \varphi_1 \cos(\omega t) + \varphi_2 \sin(\omega t),
\label{idsjbvcfrgpwuiUPIopsfd}
\ee
$ \varphi_1, \, \varphi_2 \in \R$. We have 
$
 f(t) = f_1 e^{-i\omega t} + f_2 e^{i\omega t} 
$
with $ f_1 = (\varphi_1 + i\varphi_2)/2$, $ f_2 = \overline{f_1}$,
$J=1$, $n_1=-1$, $ n_2 =1$. Applying now (\ref{QQQQ2}) for this case
with $m=0$ we get 
\be
    M(q^2) \; = \; Q^{(2)}_0 \; = \; 
e^{2i\gamma_f}
\sum_{p=0}^{\infty}\frac{(-1)^p}{(p!)^2}
\left(
\frac{4|f_1|}{2\omega}
\right)^{2p}
\; = \; 
    e^{2i\gamma_f}J_0 \left(\frac{2\varphi_0}{\omega}\right),
\label{usdshdks}
\ee
where $\varphi_0 := \sqrt{\varphi_1^2 + \varphi_2^2}$ and where $ J_0$
is the Bessel function of first kind and order zero. In this case
$
\gamma_f = \varphi_2/\omega
$.

Relation (\ref{usdshdks}) shows that condition I is not empty and that
the locus in the $(\varphi_1, \; \varphi_2)$-space of the condition
$M(q^2)=0$ (necessary for condition II) is the countable family of
circles centered at the origin with radius $ x_a \omega/2$, $ a = 1,
\,2, \ldots$, where $ x_a$ if the $a$-th zero of $ J_0$ in $ \R_+$.

One shows analogously that
\be
     Q_m \; = \;  e^{i\gamma_f}
\left(
\frac{\overline{f_1}}{|f_1|}
\right)^m
J_m \left(\frac{2|f_1|}{\omega}\right) 
\ee
and
\be
     Q^{(2)}_m \; = \;  e^{2i\gamma_f}
\left(
\frac{\overline{f_1}}{|f_1|}
\right)^m
J_m \left(\frac{4|f_1|}{\omega}\right) ,
\ee
for all $ m\in \Z$.

For $ Q^{(2)}_0 =0 $ the function $ \calQ_1$ is periodic and we have
in general
\be
    M(\calQ_1) \; = \; \frac{i}{\omega}\sum_{m\in \Z \atop m\neq 0}
    \frac{\left| Q^{(2)}_m\right|^2}{m}
\; = \;
\frac{i}{\omega}
\sum_{m=1}^{\infty}
\left(
\frac{\left| Q^{(2)}_m\right|^2 - \left| Q^{(2)}_{-m}\right|^2}{m}
\right)
\label{lkvjsxSc}
\ee

Since $|J_m(x)|=|J_{-m}(x)|$ for all $ x\in \R$, $\forall m\in \Z $,
it follows that $|Q^{(2)}_m| = |Q^{(2)}_{-m}|$, $\forall m\in \Z $.
Hence, for functions $ f$ like (\ref{idsjbvcfrgpwuiUPIopsfd})
\be
     M(\calQ_1) \; = \; 0 .
\label{condIIehvazia}
\ee

Therefore, condition II is nowhere fulfilled. For a complete solution
of the problem for functions like (\ref{idsjbvcfrgpwuiUPIopsfd}),
including the circles mentioned above, higher restrictions than that
implied by condition II are necessary.

\subsection{A Second Example}

For functions $ f$ with $ J > 1$ the situation leading to
(\ref{condIIehvazia}) is not expected in general and condition II, and
eventually others, may hold in non-empty regions of the parameter
space of $ f$.  This can be seen in the following example with $ J=2$.
Let us take
\be
    f(t) \; = \; f_1 (t) + f_2(t)
\ee
with
\bear
f_1(t) & = & f_1e^{-i\omega t} + \overline{f_1}e^{i\omega t} 
\\
f_2(t) & = & f_2e^{-i2\omega t} + \overline{f_2}e^{i2\omega t} 
\eear
$f_i\in \C$, $i=1, \, 2$. 
We have
$
 q(t) = q_1(t)q_2(t)
$, 
where
\bear
q_1(t) & := & e^{i\gamma_{f_1}}\sum_{n\in\Z}e^{in\zeta_1}
              J_n\left( \frac{2|f_1|}{\omega}\right)e^{in\omega t},
\\
q_2(t) & := & e^{i\gamma_{f_2}}\sum_{n\in\Z}e^{in\zeta_2}
              J_n\left( \frac{|f_2|}{\omega}\right)e^{in2\omega t},
\eear
with
$$
   e^{i\zeta_i} = \frac{\overline{f_i}}{|f_i|}, \qquad i=1, \, 2.
$$
It follows that
\bear
Q_m & = & e^{i(\gamma_{f_1}+\gamma_{f_2})}\sum_{k\in\Z}
          e^{i( (m-2k)\zeta_1+k\zeta_2)}
          J_{m-2k}\left( \frac{2|f_1|}{\omega}\right)
          J_{k}\left( \frac{|f_2|}{\omega}\right) ,
\\
Q_m^{(2)} & = & e^{2i(\gamma_{f_1}+\gamma_{f_2})}\sum_{k\in\Z}
                e^{i( (m-2k)\zeta_1+k\zeta_2)}
                J_{m-2k}\left( \frac{4|f_1|}{\omega}\right)
                J_{k}\left( \frac{2|f_2|}{\omega}\right) .
\eear

From this we see (using $J_{-n}(x)=(-1)^nJ_n(x)$) that
\be
\overline{Q_{-m}^{(2)}} \; = \; 
(-1)^m e^{-4i(\gamma_{f_1}+\gamma_{f_2})}
\left\{
                  e^{2i(\gamma_{f_1}+\gamma_{f_2})}\sum_{k\in\Z}
                  (-1)^k e^{i( (m-2k)\zeta_1+k\zeta_2)}
                  J_{m-2k}\left( \frac{4|f_1|}{\omega}\right)
                  J_{k}\left( \frac{2|f_2|}{\omega}\right)
\right\} .
\ee
The factor between brackets differs from $Q_{m}^{(2)} $ due to the
presence of the factor $ (-1)^k$ in the sum over $ k\in\Z$. Hence, we
should rather expect $|Q_{m}^{(2)} | \neq |Q_{-m}^{(2)} |$ in this
case, what most likely implies $M(\calQ_1) \neq 0$ for $ M(q^2)=0$,
leading to a non-empty condition II.

\newpage

\begin{appendix}

\noindent\hrulefill

\bc
  \begin{Large}
    {\bf Appendices }
  \end{Large}
\ec

\noindent\hrulefill

\section{Short Description of the Strategy Followed in \cite{qp}}
\zerarcounters
\label{ShortDescription}

For convenience of the reader we reproduce the main steps of the
strategy developed in \cite{qp} for finding a power series solution of
the generalized Riccati equation (\ref{primeiraequacaodeRiccati})
without secular terms.

As discussed in Section \ref{Introduction}, a natural proposal is to
express $g$, a particular solution of
(\ref{primeiraequacaodeRiccati}), as a formal power expansion on $
\eps$ which vanishes at $\eps=0 $. For convenience, we write this
expansion as in (\ref{expansaodeg}) where $ q(t)$ is defined in
(\ref{definicaodeq}).  This would give the desired solution, provided
the infinite sum converges.  Inserting (\ref{expansaodeg}) into
(\ref{primeiraequacaodeRiccati}) leads to
\be
  \sum_{n=1}^{\infty}
  \left(
  (q c_n) ' - i \sum_{p=1}^{n-1} q^2 c_p c_{n-p}   - 2 i  f q c_n
  \right) \eps^n 
  + i\eps^2 =0 .
\ee

Assuming that the coefficients vanish order by order
we conclude
\bear
 &  & (q c_1)' - 2 i f q c_1 = 0, \label{E1} \\
 &  & (q c_2)' - i q^2 c_1^2 - 2 i f q c_2 + i = 0,  \label{E2}\\
 &  & (q c_n)' - i \sum_{p=1}^{n-1} q^2 c_p c_{n-p} - 2 i  f q c_n = 0,
      \quad n \geq 3. 
      \label{E3}
\eear
The solutions of (\ref{E1})-(\ref{E2}) are
\bear
 & & c_1 (t) \; = \; \alpha_1 \, q(t),        \label{S1} \\
 & & c_2 (t) \; = \; 
     q(t)  \, \left[ i \int_{0}^{t} 
               \left(\alpha_1^2 q(t')^2 - q(t')^{-2}  \right) dt' + \alpha_2
          \right],        \label{S2} \\
 & &    c_n (t) \; = \; 
         q(t)  \, \left[ i \left(\sum_{p=1}^{n-1}\int_{0}^{t} 
                 c_p(t') c_{n-p}(t') \, dt' \right) + \alpha_n
              \right],  \quad \mbox{for } n\geq 3,      \label{S3} 
\eear
where the $\alpha_n$'s above, $ n = 1, \, 2, \ldots, $ are arbitrary
integration constants.

The key idea is to fix the integration constants $ \alpha_i$ in such a
way as to eliminate the constant terms from the integrands in
(\ref{S2}) and (\ref{S3}). The remaining terms involve sums of
exponentials like $ e^{in\omega t}$, $ n\neq 0$, which do not develop
secular terms when integrated, in contrast to the constant terms. For
instance, fixing $ \alpha_1 $ such that $ M(\alpha_1^2 q^2 -
q^{-2})=0$, that means, $\alpha_1^2 = M(q^{-2})/M(q^2) $, 
prevents secular terms in (\ref{S2}).

As shown in \cite{qp} this procedure can be implemented in all orders,
fixing all constants $ \alpha_i$ and preventing secular terms in all
functions $ c_n(t)$. In case I, relations
(\ref{CFourier1})-(\ref{CFouriern}) represent precisely relations
(\ref{S1})-(\ref{S3}) in Fourier space with the integration constants
fixed as explained above. Case II is analogous.

\section{The Decay of the Fourier Coefficients of $q$ and $q^2$}
\zerarcounters
\label{apendice1}

To prove our main results on the Fourier coefficients of the functions
$c_n$ and $e_n$ we have to establish some results on the decay of the
Fourier coefficients of $q$ and $q^2$.

We write the Fourier series (\ref{primeiradecomposicaodeFourierdef})
of $ f$ in the form\footnote{As above, here we adopt $F_0 =0$.}
$$
   f(t) \; = \;  \sum_{ n \in \Z \atop n \neq 0}
   F_{n} e^{i n \omega t},
$$
with $ \overline{F_{n}} = F_{-n}$, since $f$ is real.  In order to
simplify our analysis we will consider here the case where the sum
above is a finite sum. This situation is physically more realistic
anyway.

By assumption, the set of integers $\{ n\in\Z | \; F_n \neq 0\}$ is a finite
set and, by the condition that $f$ is real and $F_0 =0$, it contains
an even number of elements, say $2J$ with $ J\geq 1$.  Let us write
this set of integers as $ \{ n_1, \ldots, n_{2J}\}$ and write
\be 
    f(t) = \sum_{a=1}^{2J} f_a e^{i n_a \omega t}, 
\label{efugbyhwvpe}
\ee 
with the convention that $n_a = - n_{2J - a+1}$, for all $1 \leq a \leq J$,
with $f_a \equiv F_{n_a}$. Clearly
$\overline{f_a} = f_{2J -a +1}$, $1\leq a \leq J$.

A simple computation (see \cite{qp}) now shows that $q$ has a Fourier
decomposition as in (\ref{representacaodeFourierdeq}) with 
\be 
Q_{m} \; = \; e^{i\gamma_f} \sum_{p_1 , \, \ldots ,
     \, p_{2J} = 0 }^{\infty} \delta\left(P,\; m \right)
   \prod_{a=1}^{2J} \ \left[ \frac{1}{p_a !}  \left(
       \frac{f_a}{ n_a  \omega } \right)^{p_a} \right] ,
\label{QQQQ}
\ee
where
\be
     P \; \equiv\; P(p_1, \ldots , p_{2J}, n_1 , \ldots, n_{2J} )
      \; := \;
              \sum_{b=1}^{2J}p_b n_b  \in \Z,
\ee
and where
\be
   \gamma_f \; := \;  i\sum_{a=1}^{2J} \frac{f_a}{n_a \omega } .
\label{definicaodegamaf}
\ee
As one easily sees, $\gamma_f \in \R$. Above $\delta\left(P,\;
  m\right)$ is the Kr\"onecker delta:
$$
   \delta\left(P,\; m\right) \; := \;
       \left\{
       \begin{array}{ll}   
           1, & \mbox{if } P = m, \\
           0, & \mbox{else.}
         \end{array}
       \right.
$$

Since the function $ q^2$ is obtained from $ q$ by replacing $f\to 2f$
we have from (\ref{QQQQ})
\be 
Q_{m}^{(2)} \; = \; e^{2i\gamma_f} \sum_{p_1 , \, \ldots ,
     \, p_{2J} = 0 }^{\infty} \delta\left(P,\; m \right)
   \prod_{a=1}^{2J} \ \left[ \frac{1}{p_a !}  \left(
       \frac{2f_a}{ n_a  \omega } \right)^{p_a} \right] ,
\label{QQQQ2}
\ee
where $ Q_{m}^{(2)}$ are the Fourier coefficients of $ q^2$. 
The coefficients $ Q_m$ and $Q_{m}^{(2)} $ can also be expressed in
terms of Bessel functions of the first kind and integer order. See
Section \ref{hierarquias} for some examples.

As in \cite{qp}, define
$$ 
    \varphi \; := \; \max_{1 \leq a \leq 2J}\; 
              \left| \frac{f_a}{n_a\omega} \right|.
$$
and
$$
   \calN \; := \; \sum_{b=1}^{2J} |n_b| .
$$
Notice that, since the $n_b$'s are fixed by the choice of 
$f$, $ \calN $ is non-zero.

The following important bounds have been proven
in \cite{qp}, Appendix D:
\be
  |Q_{m}| \; \leq \; 
  \left( 2J e^{(2J-1)\varphi}\right) \; \frac{\varphi^{\lceil \calN^{-1}
  |m|\rceil}}{\lceil \calN^{-1} |m|\rceil!}\;
  \left(1 - \frac{\varphi}{\lceil \calN^{-1} |m|\rceil+1}\right)^{-1},
\label{desigualdadeforteparaQm}
\ee 
and
\be
  |Q_{m}^{(2)}| \; \leq \; 
  \left( 2J e^{(2J-1)2\varphi}\right) \; \frac{(2\varphi)^{\lceil \calN^{-1}
  |m|\rceil}}{\lceil \calN^{-1} |m|\rceil!}\;
  \left(1 - \frac{2\varphi}{\lceil \calN^{-1} |m|\rceil+1}\right)^{-1},
\label{desigualdadeforteparaQm2}
\ee 
for all $ m$ with $ \lceil \calN^{-1} |m|\rceil+1 > 2\varphi$. Above
$\lceil x \rceil$ is the lowest integer larger than or equal to $x$.

In \cite{qp} we derived from (\ref{desigualdadeforteparaQm}) a simple
exponential bound for $|Q_{m}|$, namely,
\be
     |Q_{m}| \; \leq \; \calQ\, e^{-\chi |m|},
\label{oldbound}
\ee
where $ \calQ$ and $ \chi$ are some positive constants. For the
purposes of this paper a sharper bound than (\ref{oldbound}) is needed
and we have to study relation (\ref{desigualdadeforteparaQm}) more
carefully.  The result is expressed in Proposition
\ref{proposicaosobreodecaimentodeQm} whose proof we present now.

\noindent {\bf Proof of Proposition
  \ref{proposicaosobreodecaimentodeQm}.}  Let us consider first the
coefficients $ Q_m$.  Due to the dominating factor $\lceil \calN^{-1}
|m|\rceil! $, one has
$$
    \lim_{|m|\to\infty} \frac{\novomod{m}^2}{e^{-\chi |m|}}
  \frac{\varphi^{\lceil \calN^{-1}
  |m|\rceil}}{\lceil \calN^{-1} |m|\rceil!} \; = \; 0 .
$$
for any constant $ \chi >0$. Hence, one can choose a constant $ M_1
>0$ depending on $\chi$ such that 
$$ 
\frac{\varphi^{\lceil \calN^{-1}
  |m|\rceil}}{\lceil \calN^{-1} |m|\rceil!}
 \leq 
M_1\, \frac{e^{-\chi |m|}}{\novomod{m}^2} 
$$
for all $ m\in \Z$.  Therefore, there exists a positive constant $
\calQ_1 >0$ (depending on $ \chi$) such that $|Q_m| \leq \calQ_1
\novomod{m}^{-2} e^{-\chi |m|} $ for all $ m\in \Z$.  For $Q_m^{(2)}$
we proceed in the same way and get the bound $|Q_m^{(2)}| \leq \calQ_2
\novomod{m}^{-2} e^{-\chi |m|} $ for all $ m\in \Z$. In (\ref{RR1})
and (\ref{RR2}) we adopt $ \calQ =\max\{\calQ_1, \; \calQ_2\}$.  $
\EndofProof$

{\it Remark.} The proof of Proposition
\ref{proposicaosobreodecaimentodeQm} shows that we have also 
sharper bounds like
$$
      |Q_m| \; \leq \; \calQ_k\,  \frac{e^{-\chi |m|}}{\novomod{m}^k} 
$$
for any $ k\in\N$. For the purposes of the present paper it was enough
to consider $ k=2$. 

\section{Bounds on Convolutions}
\zerarcounters
\label{apendice2}

Here we will prove Lemma \ref{lemaauxilialsobredecaimentodasconvolucoes}.
Consider for $ \chi > 0$ and $ m \in \Z$
\be
   \calB (m)  \equiv \calB(m, \, \chi) \; := \;  \sum_{n \in \Z }
\frac{e^{-\chi(|m-n|+|n|)}}{\novomod{m-n}^2\; \novomod{n}^2  } .
\label{oiuwyvbeiuvwb}
\ee
First notice that $ \calB(m) = \calB(-m)$ for all $ m\in \Z$.
Choosing $ B_0$ to be such that 
$$
   B_0 \geq \sum_{n \in \Z }
\frac{e^{-2\chi|n| }}{\novomod{n}^4  }
$$
the statement of the lemma becomes trivially correct for $ m=0$.
Hence, it is enough to consider the case where $ m >0$.

In (\ref{oiuwyvbeiuvwb}), the sum over all $ n\in \N$ can be split
into three sums: 
\be
\calB (m) \; = \;
e^{-\chi m}\sum_{n=-\infty}^{-1}\frac{e^{2\chi n}}{(m-n)^2 n^2}
+
e^{-\chi m}\sum_{n=0}^{m}\frac{1}{\novomod{m-n}^2 \; \novomod{n}^2}
+
e^{\chi m}\sum_{n=m+1}^{\infty}\frac{e^{-2\chi n}}{(m-n)^2 n^2} 
\ee
In the first sum above we perform the change of variables $ n\to -n$
and in the third sum we perform the change of variables $ n\to n+m$. 
The result is
\be
\calB (m) \; = \;
e^{-\chi m}
\left(
2\sum_{n=1}^{\infty}\frac{e^{-2\chi n}}{(m+n)^2 n^2}
+
\sum_{n=0}^{m}\frac{1}{\novomod{m-n}^2 \; \novomod{n}^2}
\right)
\label{erjhvne}
\ee

Now we will study separately each of the sums in (\ref{erjhvne}). Since
for $ n\geq 1$ one has $ m+n \geq \; \novomod{m}$ one has for the first sum
\be
\sum_{n=1}^{\infty}\frac{e^{-2\chi n}}{(m+n)^2 n^2}
\leq
\frac{B_1}{\novomod{m}^2}
\ee
where
$
\displaystyle
B_1 := \sum_{n=1}^{\infty}\frac{e^{-2\chi n}}{ n^2}
$. 

The second sum in (\ref{erjhvne}) is a little more involving. 
We have
\be
\sum_{n=0}^{m}\frac{1}{\novomod{m-n}^2 \; \novomod{n}^2}
=
\sum_{n=0}^{\meufloor{m/2}}\frac{1}{\novomod{m-n}^2 \; \novomod{n}^2}
+
\sum_{n=\meufloor{m/2}+1}^{m}\frac{1}{\novomod{m-n}^2 \; \novomod{n}^2}
\label{Tusgephrg}
\ee
For the first sum in the right hand side of (\ref{Tusgephrg}) we have
$
  \novomod{m-n} \; \geq m-n \geq m - \meufloor{m/2} \geq m/2
$. 
For the second sum in the right hand side of (\ref{Tusgephrg}) we have
$
  n \geq \meufloor{m/2}+1 \geq m/2
$. Hence, for $ m>0$, 
\bear
\sum_{n=0}^{m}\frac{1}{\novomod{m-n}^2 \; \novomod{n}^2}
 & \leq &
\left( \frac{2}{m} \right)^2
\left[
\sum_{n=0}^{\meufloor{m/2}}\frac{1}{\novomod{n}^2}
+
\sum_{n=\meufloor{m/2}+1}^{m}\frac{1}{\novomod{m-n}^2 }
\right] \nonumber \\
& \leq &
2\left( \frac{2}{\novomod{m}} \right)^2
\sum_{n=0}^{\infty}\frac{1}{\novomod{n}^2}
\eear

Therefore, choosing
\be
     B_0 = 2B_1 + 8 \sum_{n=0}^{\infty}\frac{1}{\novomod{n}^2} 
\ee
the lemma is proven. \EndofProof

The proof of this lemma has the following proposition as corollary,
generalizing Lemma \ref{lemaauxilialsobredecaimentodasconvolucoes}:
\begin{proposition}
For $\chi >0$, $ k\in \N$, $ k\geq 2$, let
\be
   \calB_k (m) \; := \;  \sum_{n \in \Z }
\frac{e^{-\chi(|m-n|+|n|)}}{\novomod{m-n}^k\; \novomod{n}^k  } . 
\ee
Then, there exists a constant $ B_{0, \; k}$, depending eventually on $
k$, such that 
\be
 \calB_k (m) \; \leq \; B_{0, \; k}\; \; \frac{e^{-\chi|m|}}{\novomod{m}^k}
\ee
for all $ m\in \Z$. $\EndofStatement$
\end{proposition}

\section{Catalan Numbers. Bounds on the Constants $ K_n$}
\zerarcounters
\label{ProvadaProposicaotipoCatalan}

Here we will prove Theorem \ref{ProposicaotipoCatalan}.  Let us start
recalling that we have chosen $ K_1 = K_2 = \calC_1$ for some constant
$ \calC_1$ which, in turn, can be chosen without loss to be larger
than or equal to $1$. The proof of Theorem \ref{ProposicaotipoCatalan}
will be presented on four steps.

\noindent{\bf Step 1.} In this step we show that the sequence $K_n$, defined in
(\ref{relacaorecursivadosKns}), is an increasing sequence.

First notice that $K_3 = \calC_2 (2K_1K_2 + (K_2)^2)=3\calC_2 (K_2)^2 $.
Since $ K_1=K_2 \geq 1$ and $ \calC_2 \geq 1$, we have $ K_1 = K_2 <
K_3$.

Let us now suppose that 
\be
   K_1 = K_2 < K_3 < \cdots < K_n
\label{ifugpbveev}
\ee
for some $ n \geq 3$. We will show that $ K_{n+1} > K_n$. 
We have
\bear 
   K_{n+1}-K_n & = &  
\calC_2 
\left[
\sum_{p=1}^{n} K_p K_{n-p+1} 
+
\sum_{p=2}^{n} K_p K_{n-p+2}
-\sum_{p=1}^{n-1} K_p K_{n-p} 
-
\sum_{p=2}^{n-1} K_p K_{n-p+1}
\right] 
\nonumber \\
 & = &  
\calC_2 
\left[
2K_1K_n +
\sum_{p=2}^{n} K_p K_{n-p+2}
-\sum_{p=1}^{n-1} K_p K_{n-p} 
\right] 
\nonumber \\
 & = &
\calC_2 
\left[
2K_1K_n +
(K_2 K_n -K_{n-2} K_1)
+
(K_3 -K_1)K_{n-1}+\cdots + (K_n - K_{n-2})K_2
\right] 
\nonumber \\
 & = &
\calC_2 
\left[
2K_1K_n +
(K_n -K_{n-2})K_1
+
(K_3 -K_1)K_{n-1}+\cdots + (K_n - K_{n-2})K_2
\right] ,
\nonumber
\eear
where in the last equality we used $K_1 =K_2$. Now, from hypothesis
(\ref{ifugpbveev}) we conclude that $ K_{n+1} > K_n$, thus proving
that $ K_n$ is an increasing sequence.

\noindent{\bf Step 2.} Here we show that the sequence $K_n$ defined in 
(\ref{relacaorecursivadosKns}) satisfies
\be
K_n \; \leq \; 
3 \calC_2 \sum_{p=2}^{n-1} K_p K_{n-p+1}
\label{desigualdademestra}
\ee
for all $n \geq 3 $. 

We have already shown that $K_3 =3 \calC_2 (K_2)^2$. Hence,
(\ref{desigualdademestra}) is obeyed for $ n=3$.

Assume now that (\ref{desigualdademestra}) is satisfied for all $ K_p$
with $p\in \{1 , \ldots , n-1\}$, for some $ n\geq 4$. We will show 
that it is also satisfied for $K_n$. In fact, we have from
(\ref{relacaorecursivadosKns})
\be
K_n \; = \; 
\calC_2 \left[
K_1 K_{n-1} +
K_2 (K_{n-2}+K_{n-1}) +
K_3 (K_{n-3}+K_{n-2}) + \cdots +
K_{n-1} (K_1+K_2)
\right] .
\ee
From this and from the fact proven in step 1 that the sequence $ K_n$
is increasing, it follows that
\be
K_n \; \leq \; 
\calC_2 \left[
K_1 K_{n-1} +
2\left( 
K_2 K_{n-1} +
K_3 K_{n-2} +   \cdots +
K_{n-1} K_2 
\right)
\right]
\label{ldiufhgnpewr}
\ee
Now, using the obvious relation
$$
K_1 K_{n-1}
=
K_2 K_{n-1}
\leq
\left( 
K_2 K_{n-1} +
K_3 K_{n-2} +  \cdots +
K_{n-1} K_2 
\right)
$$
we get finally from (\ref{ldiufhgnpewr})
\be
K_n \; \leq \; 
3 \calC_2 \left[
K_2 K_{n-1} +
K_3 K_{n-2} +  \cdots +
K_{n-1} K_2 
\right]
\; = \; 
3 \calC_2
\sum_{p=2}^{n-1} K_p K_{n-p+1} ,
\ee
thus proving (\ref{desigualdademestra}).

\noindent{\bf Step 3.} Here we will prove the following statement. Let $ L_n$
be defined as the sequence such that $L_1 = L_2 =K_1 =K_2=\calC_1$ and
\be
    L_n \; = \; 3 \calC_2
\sum_{p=2}^{n-1} L_p L_{n-p+1}.
\label{definicaodasequanciaLn}
\ee
Then, one has 
\be 
       K_n \leq L_n , \qquad \forall n\in \N  .
\label{seiufhnv}
\ee 

First notice that $ K_3 = 3\calC_2 (K_1)^2 = 3\calC_2 (L_1)^2 =L_3 $.
Hence, (\ref{seiufhnv}) is valid for $n\in \{1, \; 2, \; 3\}$.  Now
suppose $K_p \leq L_p $ for all $p \in \{1, \ldots , n-1\}$ for some 
$n \geq 4$. One has from (\ref{desigualdademestra})
\be
K_n \; \leq \; 
3 \calC_2 \sum_{p=2}^{n-1} K_p K_{n-p+1}
\; \leq \; 
3 \calC_2 \sum_{p=2}^{n-1} L_p L_{n-p+1}   
\; = \; 
L_n ,
\ee
thus proving (\ref{seiufhnv}). 

\noindent{\bf Step 4.} Consider the sequence $ {\mathbf c}_n$ defined as follows:
$ {\mathbf c}_1 = {\mathbf c}_2 =1$ and
\be
  {\mathbf c}_n \; = \; \sum_{p=2}^{n-1} {\mathbf c}_p {\mathbf c}_{n-p+1} 
\label{eoinvw}
\ee  
for $ n\geq 3$. The so defined numbers ${\mathbf c}_n$ are called
``Catalan numbers'', after the mathematician Eug\`ene C.
Catalan. The Catalan numbers arise in several combinatorial problems
(for a historical account with proofs, see \cite{Dorrie}) and can
be expressed in a closed form as
\be
     {\mathbf c}_n \; = \; \frac{(2n-4)!}{(n-1)!(n-2)!} , \qquad n\geq 2 .
\ee
(see, f.i, \cite{Dorrie} or \cite{GrahamKnuthPatashnik}). 
Using Stirling's formula we get the following asymptotic behaviour
for the Catalan numbers:
\be
  {\mathbf c}_n \; \approx \; \frac{1}{16\sqrt{\pi}}\;\frac{4^n}{n^{3/2}} ,
      \qquad n \mbox{ large.}
\ee 

The existence of a connection between the Catalan numbers and the
sequence $ L_n$ defined above is evident. Two distinctions are the
factor $ 3 \calC_2$ appearing in (\ref{definicaodasequanciaLn}) and
the fact that $L_1=L_2 = \calC_1$ is not necessarily equal to $ 1$.
Nevertheless, using the definition of the Catalan numbers in
(\ref{eoinvw}), it is easy to prove the following closed expression
for the numbers $ L_n$:
\be
       L_n \;\; = \;\; (\calC_1)^{n-1}\;(3\calC_2)^{n-2}\;
                \frac{(2n-4)!}{(n-1)!(n-2)!} , \qquad n\geq 2 .
\ee 
We omit the proof here. Hence, the following asymptotic behaviour 
can be established:
\be
L_n \; \approx \;
\frac{1}{144 \calC_1\calC_2^2\sqrt{\pi}}\;\frac{(12\calC_1\calC_2
  )^n}{n^{3/2}} , \qquad n \mbox{ large.}
\ee 

From the inequality $ K_n \leq L_n$, proven in step 3, it follows that
$K_n \leq K_0 (12\calC_1\calC_2)^n$ for some constant $ K_0 >0$, for
all $ n\in \N$.  Theorem \ref{ProposicaotipoCatalan} is now proven.
$\EndofProof$

\end{appendix}

\newpage

\noindent\hrulefill

\noindent{\bf Acknowledgements. } {\it I am very indebted to Walter
F. Wreszinski for enthusiastically supporting this work and for many
important suggestions. I am also grateful to C\'esar R. de Oliveira
for asking the right questions.}

\vfill

\noindent\hrulefill


\end{document}